\documentclass{SciPost}

\hypersetup{
    colorlinks,
    linkcolor={red!50!black},
    citecolor={blue!50!black},
    urlcolor={blue!80!black}
}

\usepackage[bitstream-charter]{mathdesign}
\DeclareSymbolFont{usualmathcal}{OMS}{cmsy}{m}{n}
\DeclareSymbolFontAlphabet{\mathcal}{usualmathcal}

\fancypagestyle{SPstyle}{
\fancyhf{}
\lhead{\colorbox{scipostblue}{\bf \color{white} ~SciPost Physics }}
\rhead{{\bf \color{scipostdeepblue} ~Submission }}

\fancyfoot[C]{\textbf{\thepage}}
}

\usepackage{graphicx}
\usepackage{dcolumn}
\usepackage{bm}
\usepackage{multirow}
\usepackage{array}
\usepackage{diagbox}

\usepackage{comment}
\usepackage{color}
\usepackage{physics}
\usepackage[table]{xcolor}

\DeclareMathOperator*{\To}{\to}

\newcommand{\up}{\uparrow}
\newcommand{\down}{\downarrow}
\renewcommand{\k}{{\bf k}}
\newcommand{\p}{{\bf p}}

\newcommand{\q}{{\bf q}}
\newcommand{\K}{{\bf K}}

\newcommand{\0}{{\bf 0}}

\newcommand{\nn}{\nonumber}
\newcommand{\beq}{\begin{equation}}
\newcommand{\eeq}{\end{equation}}

\newcommand{\fs}{\ket{\mathrm{FS}}}

\newcommand{\new}[1]{{#1}}

\begin{document}

\pagestyle{SPstyle}

\begin{center}{\Large \textbf{\color{scipostdeepblue}{
Role of impurity statistics and medium constraints in polaron-polaron interactions\\
}}}\end{center}

\begin{center}\textbf{
Jesper Levinsen\textsuperscript{1}, 
Francesca Maria Marchetti\textsuperscript{2,3},
Olivier Bleu\textsuperscript{1,4} and
Meera M. Parish\textsuperscript{1}
}\end{center}

\begin{center}
{\bf 1} School of Physics and Astronomy, Monash University, Victoria 3800, Australia
\\
{\bf 2} Departamento de F\'isica Te\'orica de la Materia
  Condensada, Universidad
  Aut\'onoma de Madrid, Madrid 28049, Spain
\\
{\bf 3} Condensed Matter Physics Center (IFIMAC), Universidad Autónoma de Madrid, 28049 Madrid, Spain
\\
{\bf 4} Institut f{\"u}r Theoretische Physik, Universit{\"a}t Heidelberg, Philosophenweg 19, 69120 Heidelberg, Germany
\end{center}

\section*{\color{scipostdeepblue}{Abstract}}
\textbf{\boldmath{
We consider the behavior of a small density of mobile impurities (polarons) immersed in a quantum gas, a generic scenario that can be realized in cold atomic gases, liquid helium mixtures, and doped semiconductors. We present a unified theoretical framework for understanding polaron quasiparticles beyond the single-impurity limit, and we identify two key factors that control the polaron-polaron interactions: (i) the statistics of the impurities, including whether or not they are degenerate, and (ii) the constraints on the medium response, i.e., whether the medium density or chemical potential is held fixed. By constructing wave functions for two bosonic, fermionic, or distinguishable impurities immersed in a Bose or Fermi gas, we derive rigorous results for the polaron interactions in the limit of weak impurity-medium coupling. We furthermore obtain an exact relationship between the polaron interactions at fixed medium density and at fixed chemical potential, a result which is valid for arbitrary interaction strength. Our work provides an important guide for understanding experiments, and it acts as a starting point for future strong-coupling theories of polaron interactions that capture all of the effects identified in this work.
}}

\vspace{\baselineskip}

\noindent\textcolor{white!90!black}{%
\fbox{\parbox{0.975\linewidth}{%
\textcolor{white!40!black}{\begin{tabular}{lr}%
  \begin{minipage}{0.6\textwidth}%
    {\small Copyright attribution to authors. \newline
    This work is a submission to SciPost Physics. \newline
    License information to appear upon publication. \newline
    Publication information to appear upon publication.}
  \end{minipage} & \begin{minipage}{0.4\textwidth}
    {\small Received Date \newline Accepted Date \newline Published Date}%
  \end{minipage}
\end{tabular}}
}}
}


\vspace{10pt}
\noindent\rule{\textwidth}{1pt}
\tableofcontents
\noindent\rule{\textwidth}{1pt}
\vspace{10pt}

\section{Introduction}

The complexity of strongly interacting quantum systems consisting of many particles would, at first glance, appear to preclude their theoretical description. It was therefore a simplification of both great conceptual and practical importance when Landau realized that the properties of a given quantum system can be captured by a dilute collection of its elementary excitations, so-called quasiparticles~\cite{Landau1957}. Here, the quasiparticles are described by particle-like properties such as their mass, charge, and energy, as well as their mutual interactions. The quasiparticle concept has had profound implications for our understanding of a range of different systems, from liquid Helium~\cite{Bardeen1966,Bardeen1967,Saam1969}---the original quantum fluid---to dilute vapors of ultracold atoms~\cite{RMP2008Many,Giorgini2008}, and even to the description of neutron stars~\cite{Baym1971,Kobyakov2016}. It has furthermore enabled technological progress, most prominently the semiconductor technology underpinning the information age.

One of the cleanest realizations of a quasiparticle is that of a single mobile impurity particle immersed in a well-understood quantum degenerate medium such as an ideal Fermi gas or a Bose-Einstein condensate (BEC). Recent years have seen an explosion in experiments investigating such so-called Fermi and Bose polaron quasiparticles, both in the context of ultracold atoms~\cite{Schirotzek2009,Nascimbene2009,Kohstall2012,Catani2012,Koschorreck2012,Zhang2012,Wenz2013,Ong2015,Cetina2015,Cetina2016,Hu2016,Jorgensen2016,Scazza2017,Mukherjee2017,Yan2019a,Oppong2019,Yan2019,Ness2020,Adlong2020,Fritsche2021,Skou2021,Cayla2023,Baroni2024,Vivanco2025,Etrych2025} and in atomically thin semiconductors~\cite{Takemura_NaturePhys2014,Sidler_NatPhys_2017,Tan_PRX20,Wagner2020,Liu2021,Muir2022,Tan_PRX2023,Huang2023,Ni2025}. Here, the focus has been on scenarios where the underlying impurity-medium interaction is strong and attractive, such as in the vicinity of an impurity-medium two-body bound state, where the resonantly enhanced dressing of impurities by excitations of the medium leads to clear quasiparticle signatures. In this case, the spectrum of the impurity generically features two branches corresponding to attractive and repulsive polarons, whose energies are either lowered or increased due to the interaction with the medium. The experimental progress has been mirrored by significant theoretical advances in modeling polarons, such that most properties of single polarons are now well characterized. For recent reviews of polaron quasiparticles, we refer the reader to Refs.~\cite{Chevy2010,Massignan_RPP2014,Scazza2022,Parish2025Varenna,massignan2025polaronsatomicgasestwodimensional,Grusdt2025}.

There is still, however, ongoing debate about the behavior beyond the single-polaron problem, with different theories even predicting different signs of the medium-induced quasiparticle interaction. In the context of dilute solutions of $^3$He in $^4$He, $^3$He atoms with different spins have been found to have an attractive induced interaction~\cite{Bardeen1966,Bardeen1967,Saam1969}. By contrast, recent theories based on the exchange of thermal impurities~\cite{YuPethick2012,Camacho2018,Paredes2024} have predicted attractive and repulsive interactions between bosonic and fermionic impurities, respectively, and (due to the lack of exchange) the absence of mediated interactions between distinguishable impurities, which appears to contradict the well-established description of liquid Helium mixtures. Most recently, some of us have shown~\cite{levinsen2025} that the medium can even induce a repulsion between bosonic impurities immersed in a Bose-Einstein condensate, and that this is the leading-order effect when the impurities are degenerate such that there is no exchange.

Experimentally, interactions between polaron quasiparticles are also under intense investigation. Again, the sign of the quasiparticle interactions appears inconsistent across different platforms. A recent experiment~\cite{Baroni2024} on bosonic and fermionic impurities in an ultracold atomic Fermi gas observed attractive (repulsive) quasiparticle interactions between bosonic (fermionic) impurities, respectively, in agreement with theories of exchange. On the other hand, recent two-dimensional (2D) semiconductor experiments with (bosonic) exciton or exciton-polariton impurities immersed in either an electron gas~\cite{Sidler_NatPhys_2017,Tan_PRX20,Muir2022,Ni2025} or an exciton-polariton coherent state~\cite{Tan_PRX2023} have found that attractive polarons interact repulsively, rather than attractively. 

\begin{figure}[t]
\centering
\includegraphics[width=.75\columnwidth]{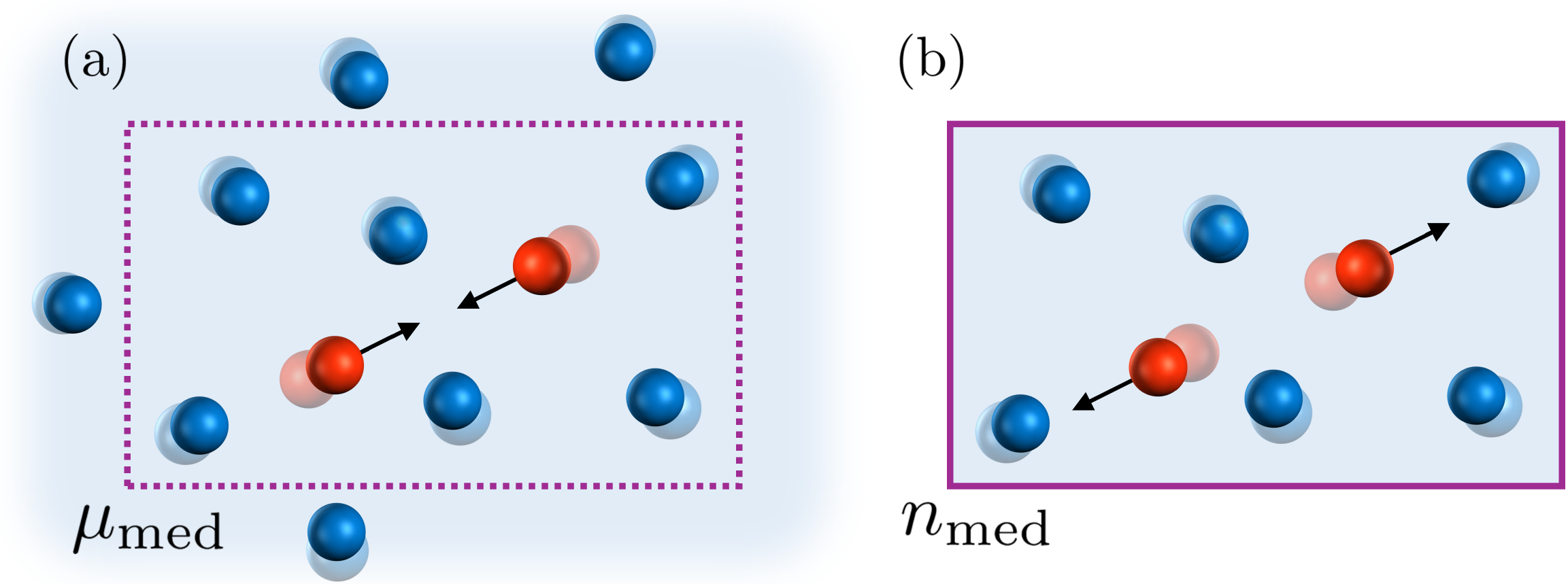}
\caption{Sketch of the different constraints on the medium (blue particles): either its chemical potential (a) or its density (b) is fixed. These can impact the strength of the interactions between polarons. For instance, for identical and degenerate bosonic impurities (red), the interactions can be attractive at fixed chemical potential and repulsive at fixed density.}
\label{fig:sketch}
\end{figure}

Here we introduce a unified framework based on well-established single-polaron theories that enables us to explain both the apparent discrepancies between theories as well as the signs of the quasiparticle interactions in the above-mentioned experiments. We identify two key factors that need to be carefully accounted for when comparing different theories and different measurements of polaron-polaron interactions across  distinct experimental platforms such as ultracold atomic gases and 2D semiconductors. The first is that the medium-induced interactions depend strongly on the constraints that are imposed on the medium's response (Fig.~\ref{fig:sketch}), i.e., on whether the medium density or chemical potential is kept fixed. As an extreme example, in the case of fermionic impurities it is already known that the polaron interactions vanish in a grand canonical description that involves chemical potentials instead of densities~\cite{Mora2010}. Similarly, non-degenerate bosonic impurities at fixed medium chemical potential can instead display a statistical enhancement of their interactions~\cite{levinsen2025}.  We show that the polaron-polaron interactions under different medium constraints obey an exact thermodynamic relation that is valid for arbitrary interaction  strengths in thermal equilibrium.

The second factor that determines the polaron-polaron interactions is the statistics of the impurities themselves. While this point has been recognized in previous theories of bosonic and fermionic impurities~\cite{YuPethick2012,Camacho2018,Scazza2022}, a crucial aspect that was overlooked was that medium-induced interactions can still exist even in the absence of exchange processes. In particular, when bosonic impurities have the \textit{same} momentum, there is no concept of exchange and one instead has medium-enhanced repulsion~\cite{levinsen2025} rather than the attraction predicted for non-degenerate impurities~\cite{YuPethick2012,Camacho2018}. We furthermore show that a similar medium-enhanced interaction occurs between distinguishable impurities.

Our theoretical framework is based on a wave-function approach that allows us to calculate the polaron-polaron interaction energy shift between two impurities of arbitrary statistics immersed in either a weakly interacting BEC or an ideal Fermi gas. Specifically, we perform a controlled perturbative expansion in the impurity-medium interaction strength up to second order. However, while we focus on the limit of weak interactions, we can straightforwardly extend our approach to describe polaron-polaron bound states (bipolarons) and other non-perturbative phenomena. We thus expect our work to form a convenient starting point for future strong-coupling theories of polaron-polaron interactions.

The manuscript is organized as follows. Section~\ref{sec:summary} provides a detailed summary of our results, and applies generally to quantum mixtures, independently of the particular realization. Section~\ref{sec:ensemble} discusses the role of constraints imposed on the medium response. In Sections~\ref{sec:bosepolaron} and \ref{sec:fermipolaron} we explore Bose and Fermi polarons, respectively, as well as the associated polaron-polaron interactions for two impurities of arbitrary statistics. In Section~\ref{sec:conc} we conclude and provide an outlook.

\section{Summary of results}
\label{sec:summary}

\begin{table*}[t]
    \centering
    \renewcommand{\arraystretch}{1.1} 
    \setlength{\tabcolsep}{10pt} 
\begin{tabular}{|c||c|c|c|c|}
    \hline
    \multirow{3}{*}{\bfseries\diagbox[width=\dimexpr 0.17\textwidth+.8\tabcolsep\relax, height=1.58cm]{\hspace{-2mm}Constraint}{{Impurities\hspace{-2mm}}}}
    & Fermions & \multicolumn{2}{c|}{Bosons} & Distinguishable \\
    \noalign{\vskip-0pt}
    & \hspace{2.3cm} & \hspace{2.3cm} & \hspace{2.3cm} & \hspace{2.3cm} \\
    \noalign{\vskip0pt}
    & & $\p_1\neq \p_2$ & $\p_1= \p_2$ & \\
    \hline \hline
    \rule{0pt}{3.3ex}Fixed $n_\mathrm{med}$ 
      & \cellcolor{blue!20}\makebox[2.3cm][c]{+} 
      & \cellcolor{orange!20}\makebox[2.3cm][c]{$-$} 
      & \cellcolor{red!30}\makebox[2.3cm][c]{sign($g_{\sigma\sigma}$)} 
      & \cellcolor{olive!20}\makebox[2.3cm][c]{sign($g_{\up\down}g_{m\up}g_{m\down}$)} \\
    \hline
    \rule{0pt}{3.3ex}Fixed $\mu_\mathrm{med}$ 
      & \cellcolor{cyan!20}\makebox[2.3cm][c]{0} 
      & {$-$} 
      & \cellcolor{magenta!20}\makebox[2.3cm][c]{$-$} 
      & \cellcolor{gray!20}\makebox[2.3cm][c]{$-$sign($g_{m\up}g_{m\down})$} \\
    \hline
\end{tabular}\\[8pt]
\begin{tabular}{@{\hspace{4pt}}c@{\hspace{4pt}}c@{\hspace{4pt}}c@{\hspace{4pt}}c@{\hspace{4pt}}}
\!$^{40}$K in $^6$Li \cite{Baroni2024}    \quad\cellcolor{blue!20} & 
\!$^{41}$K in $^6$Li \cite{Baroni2024}    \quad\cellcolor{orange!20} & 
\!X in e \cite{Muir2022,Ni2025}, P in e \cite{Tan_PRX20}, P in P   \cite{Tan_PRX2023} \quad\cellcolor{red!30} & 
\!X in e \cite{Muir2022,Ni2025} \hspace{-2mm}    \cellcolor{olive!20} \\
\end{tabular}
\begin{tabular}{@{\hspace{4pt}}c@{\hspace{4pt}}c@{\hspace{4pt}}c@{\hspace{4pt}}c@{\hspace{4pt}}}
\!$^{6}$Li in $^6$Li \cite{Nascimbene2010}  \hspace{2.3mm}\cellcolor{cyan!20} & 
\cellcolor{cyan!20} & 
\!$^{133}$Cs in $^6$Li \cite{DeSalvo2019}   \quad\cellcolor{magenta!20} & 
\!$^3$He in $^4$He \cite{Edwards1965,Anderson1966}  \hspace{-2mm} \cellcolor{gray!20} \\
\end{tabular}
    \caption{
    Sign of the \new{low-momentum} medium-induced \new{quasiparticle} interaction ($F_n-\new{f^{(0)}}$ or $F_\mu-\new{f^{(0)}}$) for different impurity statistics and medium constraints (i.e., fixed medium density $n_\mathrm{med}$ or chemical potential $\mu_\mathrm{med}$). The signs are obtained at second order in the medium-impurity interaction strengths $g_{m\sigma}$ and $g_{m\sigma'}$ and first order in the \new{bare} impurity-impurity interaction strength $g_{\sigma\sigma'}$, where the impurity species $\sigma$ and $\sigma'$ can be the same or distinguishable. We take the limits $T\to0$ and $n_\sigma,n_{\sigma'}\to0$, and thus the \new{momenta of the two impurities} $\p_1,\p_2\to0$. While the sign does not depend on the medium statistics, it does depend on the impurity statistics, including whether their momenta are the same (for instance if they form a condensate) or not. The color coding refers to experiments that have probed polaron-polaron interactions for various impurities immersed in various quantum media. Here, X, P, and e stand for excitons, exciton-polaritons, and electrons, respectively, in 2D semiconductors or semiconductor microcavities.
    } 
    \label{tab:signs}
\end{table*}

Before proceeding to our explicit calculations, we first summarize the rich phenomenology of the quasiparticle interactions between impurities. \new{Here, we consider two impurities species labeled by a pseudospin  $\sigma \in\{\up,\down\}$ that specifies the type of particle (e.g., a specific isotope of an atom) and/or the appropriate internal quantum numbers of that particle (e.g., the hyperfine spin). It is straightforward to generalize the results to multiple impurity species.}

\new{The starting point of our analysis is the energy density $\mathcal{E}$ of a dilute collection of impurity quasiparticles immersed in a quantum medium. To second order in the occupation numbers, this takes the form~\cite{BaymPethick1991book}
\begin{align}\label{eq:energydensity1}
    \mathcal{E}=\mathcal{E}_0+\frac1{V}\sum_{\p\sigma}E_{\mathrm{pol},\sigma}(\p)n_{\p\sigma}+\frac12\frac1{V^2}\sum_{\p_1\sigma,\p_2\sigma'}f_{\p_1\sigma,\p_2\sigma'}n_{\p_1\sigma}n_{\p_2\sigma'}\,.
\end{align}
Here, $V$ is the volume, $\mathcal{E}_0$ is the energy density of the medium in absence of impurities, $E_{\mathrm{pol},\sigma}(\p)$ is the quasiparticle energy of an impurity at momentum $\p$ and species $\sigma$, $n_{\p\sigma}$ is the quasiparticle distribution function, and $f_{\p_1\sigma,\p_2\sigma'}$ is the quasiparticle interaction between $\{\p_1,\sigma\}$ and $\{\p_2,\sigma'\}$ impurities. 

In practice, experiments are often conducted in a low-temperature regime where the occupation of impurities is, at least approximately, dominated by small momenta. We therefore take the limit of $T\to0$ where we can assume that the relevant part of the quasiparticle dispersion is quadratic 
\begin{align}
    E_{\mathrm{pol},\sigma}(\p)=E_{\mathrm{pol},\sigma}+\frac{p^2}{2m^*_{\sigma}}\,,
\end{align}
with $E_{\mathrm{pol},\sigma}$ the zero-momentum quasiparticle energy and $m^*_{\sigma}$ the effective mass. Furthermore, for small impurity momenta $\p_1$ and $\p_2$ we have an approximately constant quasiparticle interaction 
\begin{align}
    F_{n,\sigma\sigma'}=\lim_{\p_1,\p_2\to0}f_{\p_1\sigma,\p_2\sigma'}\,.
\end{align}
The energy density then takes the form
\begin{align}\label{eq:energydensity2}
    \mathcal{E}=\mathcal{E}_0+\sum_\sigma\left( E_{\mathrm{pol},\sigma}+\mathcal{E}_{\mathrm{kin},\sigma}\right)+\frac12\sum_{\sigma\sigma'}F_{n,\sigma\sigma'}n_\sigma n_{\sigma'}\,,
\end{align}
where $n_\sigma=\frac1V\sum_\p n_{\p\sigma}$ is the density of species $\sigma$. The term $\mathcal{E}_{\mathrm{kin},\sigma}$ arises from the quasiparticle dispersion in Eq.~\eqref{eq:energydensity1} and, in the limit  $T\to0$, is only non-vanishing in the case of identical fermionic impurities where it corresponds to the kinetic energy of a non-interacting Fermi gas: $\mathcal{E}_{\mathrm{kin},\sigma}=\frac35 \frac{(6\pi^2)^{2/3}}{2m^*_\sigma}n_{\sigma}^{5/3}$ (we work in units where $\hbar=1$).

In deriving the low-temperature energy density in Eq.~\eqref{eq:energydensity2}, we have glossed over two delicate points. First, we did not specify how we take the limit of $T\to0$. However, as we discuss further below, the order of limits is of utmost importance if our impurities are identical bosons, since they can then undergo a phase transition, i.e., if we take temperature $T\to0$ before taking the impurity density $n_\sigma\to0$ then the impurities will condense, whereas if we take $n_\sigma\to0$ before $T\to0$ then the impurities will remain thermal. Second, by working in the canonical ensemble, we implicitly kept the medium density fixed. However, it is possible to impose other constraints on the medium response to the presence of impurities, such as a fixed medium chemical potential.}

Table~\ref{tab:signs} illustrates how the sign of the induced \new{quasiparticle} interactions \new{indeed} depends strongly on both the impurity statistics\new{---including the possibility of having degenerate bosonic impurities---}and on the constraints imposed on how the medium can respond to a perturbation. Here, we focus on the medium-induced part of the interactions and subtract any bare interaction $\new{f^{(0)}}$ between the impurities. We furthermore focus on the perturbative regime where we only consider the lowest non-vanishing contribution in the impurity-medium interaction strength, which allows us to make rigorous statements about the signs. Importantly, the sign of the leading-order contribution does not depend on the details of the medium or even on whether we consider a 2D or a 3D geometry. Table~\ref{tab:signs} therefore applies generally to quantum mixtures with 1 or 2 dilute components, such as $^3$He-$^4$He mixtures, ultracold atomic gases with fermionic and/or bosonic atoms, excitons and electrons in 2D semiconductors, and even to systems featuring a strong coupling between light and matter, such as exciton-polaritons in semiconductor microcavities. In particular, we see that previous experiments on polaron-polaron interactions have probed a range of different scenarios, and this must be accounted for when interpreting the results.

\subsection{Bare interaction between impurities}
Throughout this paper we will focus on short-range interactions, as is appropriate for ultracold atoms or excitons in two-dimensional semiconductors. Therefore, in the absence of a medium, there are no interactions between identical fermionic impurities, but we can have interactions between either bosonic or distinguishable impurities. 

More precisely, if we have two particles (impurities) with low momenta $\p_1$ and $\p_2$ in a volume $V$, then their (bare) interaction energy will be \new{$f^{(0)}/V$}, where the coefficient \new{$f^{(0)}$} is a constant that takes the form 
\begin{equation}
    \new{f^{(0)}}= \begin{cases} 0 & \text{fermions} \\ g_{\sigma\sigma} & \text{bosons }(\p_1=\p_2) \\
    2g_{\sigma\sigma} & \text{bosons }(\p_1\neq\p_2) \\ g_{\up\down} & \text{distinguishable}
    \,.
    \end{cases}
\end{equation}
Importantly, we see that the interaction energy for two (identical) bosons depends on whether or not they have the same momenta, a consequence of the possibility of exchanging bosons with distinct momenta. This process is also crucial in the case of fermionic impurities, which naturally have distinct momenta. In this case, rather than an enhancement we have a cancellation of direct and exchange processes which results in a vanishing interaction energy shift.

In three dimensions, the interaction coefficient between two identical bosonic impurities at low momentum takes the form $g_{\sigma\sigma}=\frac{4\pi a_{\sigma\sigma}}{m_\sigma}$, with $a_{\sigma\sigma}$ and $m_\sigma $ the corresponding scattering length and mass, respectively. Likewise, the interaction coefficient of distinguishable impurities of species $\up$ and $\down$ is $g_{\up\down}=\frac{2\pi a_{\up\down}}{m_{\up\down}}$ in terms of the scattering length $a_{\up\down}$ and reduced mass $m_{\up\down}=m_\up m_{\down}/(m_\up +m_{\down})$.

\subsection{Medium constraints}

As is immediately clear from Table~\ref{tab:signs}, the quasiparticle interactions strongly depend on the constraints imposed on the medium response, namely whether the medium density $n_\mathrm{med}$ or chemical potential $\mu_\mathrm{med}$ (or, equivalently, pressure) is kept fixed. This should not come as a surprise since the polaron-polaron interactions arise from the density-density response of the medium, and it is only natural that this should be sensitive to the constraints imposed on said response. In this sense, it is similar to the specific heat in thermodynamics, which also depends on whether the volume or the pressure is kept fixed. Furthermore, like the case of specific heat, the polaron interactions under the different medium constraints can be directly related. To do so, we use thermodynamic arguments similar to those of Ref.~\cite{Bardeen1967} in the context of $^3$He impurities in superfluid $^4$He (for additional details of the calculation, see Section~\ref{sec:ensemble}). 

At fixed medium density, \new{Eq.~\eqref{eq:energydensity2} implies that the low-momentum} quasiparticle interaction between two impurities of species $\sigma$ and $\sigma'$ is
\begin{align}
    F_{n,\sigma\sigma'}=\left.\pdv{^2\new{\mathcal{E}'}}{n_\sigma\partial n_{\sigma'}}\right|_{n_\mathrm{med}}
    \,,
    \label{eq:Fn}
\end{align}
where 
we take the limit of vanishing impurity densities $n_\sigma,n_{\sigma'}$. In the absence of the medium, this definition simply yields $F_{n,\sigma\sigma'}=\new{f^{(0)}}$. \new{The prime on $\mathcal{E}$ signifies how, in the case of fermionic impurities, we must remove the single-particle 
kinetic-energy term $\sim n_\sigma^{5/3}$ appearing in Eq.~\eqref{eq:energydensity2} 
before performing the second derivative since this does not constitute an interaction between polarons.}

Similarly, at fixed medium chemical potential, we have
\begin{align}
    F_{\mu,\sigma\sigma'}=\left.\pdv{^2\new{\Omega'}}{n_\sigma\partial n_{\sigma'}}\right|_{\mu_\mathrm{med}}
    \,,
    \label{eq:Fmu}
\end{align}
where the grand potential $\Omega = \mathcal{E} - \mu_\mathrm{med} n_\mathrm{med}$, and any kinetic-energy contribution must be removed like before. Thermodynamic relations then allow us to exactly relate these two different quasiparticle interaction strengths as follows: 
\begin{align}
    F_{\mu,\sigma\sigma'}=F_{n,\sigma\sigma'}-\frac{\Delta N_\sigma\Delta N_{\sigma'}}{\mathcal{N}}\,.
    \label{eq:FmuFn}
\end{align}
Here, $\Delta N_\sigma=\left.\pdv{n_\mathrm{med}}{n_\sigma}\right|_{\mu_\mathrm{med}}$ is the number of medium particles in the polaron dressing cloud~\cite{Massignan2005} and $\mathcal{N}=\pdv{n_\mathrm{med}}{\mu_\mathrm{med}}$ is the density of states at the medium chemical potential, where  $\mathcal{N} > 0$ in order for the medium to be stable.\footnote{\new{As discussed further in Section~\ref{sec:ensemble},} in the context of liquid He mixtures, $F_{n,\sigma\sigma'}$ and $-\Delta N_\sigma\Delta N_{\sigma'}/\mathcal{N}$ are termed the direct and induced interactions, respectively~\cite{Bardeen1967}, while in papers that focus on exchange interactions at fixed medium density, such as Ref.~\cite{YuPethick2012,Camacho2018,Paredes2024}, $\left.F_n\right|_{\new{f^{(0)}}=0}$ is referred to as the mediated interaction.}

The relationship in Eq.~\eqref{eq:FmuFn} is reflected in the large variety of possible signs of the medium-induced interaction in Table~\ref{tab:signs}. For the case of identical impurities ($\sigma = \sigma'$), Eq.~\eqref{eq:FmuFn} demonstrates that the quasiparticle interactions contain an attractive contribution when one considers a fixed medium chemical potential, as depicted in Fig.~\ref{fig:sketch}. Physically, this is due to the fact that identical impurities can lower the system's energy by perturbing the medium density. The situation is even more nuanced for distinguishable impurities at fixed $\mu_\mathrm{med}$: if one impurity attracts while the other repels the medium, then $-\Delta N_{\up} \Delta N_{\down}/\mathcal{N}>0$ such that the resulting medium-induced interaction is \textit{repulsive}. 

\begin{figure}[t]
\centering
\includegraphics[width=.75\columnwidth]{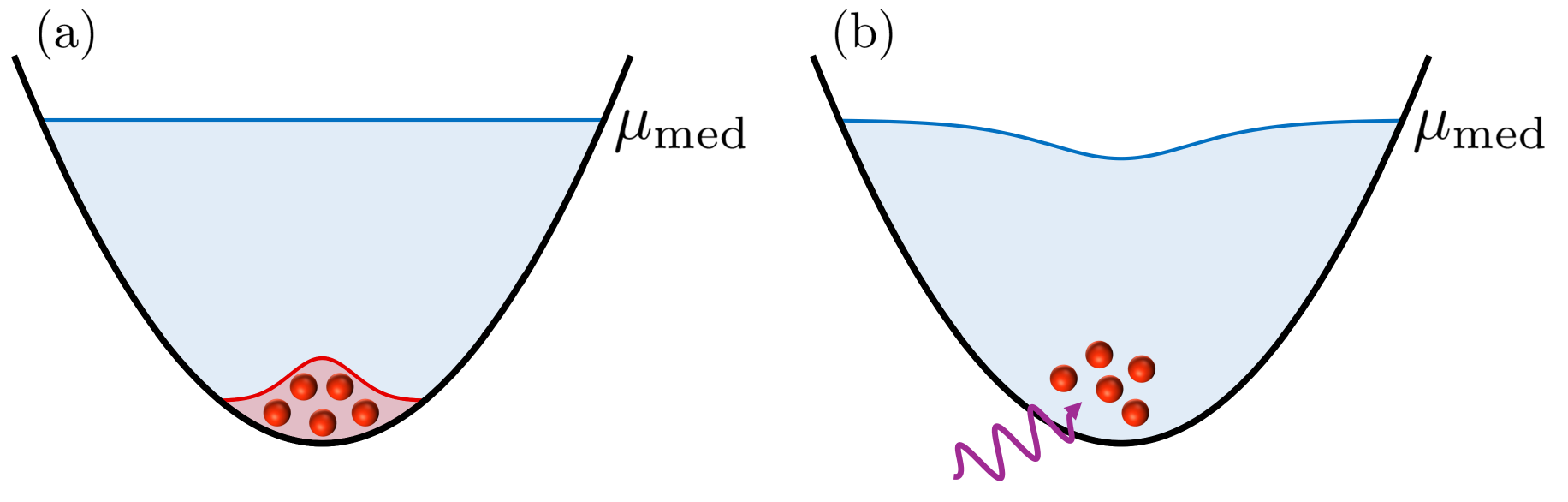}
\caption{Sketch of the different measurement protocols carried out in a trapped system. In an equilibrium measurement (a), such as one that probes the shape of the impurity cloud, the medium chemical potential is kept fixed across the trap. On the other hand, in an injection measurement (b) carried out at a time scale short compared with the (inverse) trap frequency, the medium chemical potential is deformed locally while the density is kept constant.}
\label{fig:sketch2}
\end{figure}

While the case of fixed medium density may appear to be the more natural constraint on the medium response,  we emphasize that the case of fixed chemical potential is  experimentally very relevant. In particular, it forms the basis of older theories of polaron interactions in $^3$He-$^4$He mixtures~\cite{Bardeen1967}, as we discuss further in Section~\ref{sec:helium}. Furthermore, trapped ultracold atomic gases are typically well described within the local density approximation~\cite{Giorgini2008}: Here, at any given position in the trap, the medium behaves like a uniform gas characterized by a local chemical potential $\mu_\mathrm{med}(\mathbf{r})=\mu_\mathrm{med}-V_\mathrm{trap}(\mathbf{r})$, where $\mu_\mathrm{med}$ is the chemical potential at the center of the trap and $V_\mathrm{trap}(\mathbf{r})$ the trapping potential. Therefore, when impurities are introduced at the center of the trap, one can naturally realize the scenario of fixed medium chemical potential (see Fig.~\ref{fig:sketch2}), as we discuss further below.

\subsection{Impurity statistics and degeneracy}

The effective polaron interactions also depend strongly on the statistics of the polaron quasiparticles. Previous works~\cite{Yu2010,YuPethick2012,Camacho2018,Paredes2024} have argued that for indistinguishable impurities where $\sigma=\sigma'$, these quasiparticle interactions are dominated by the exchange term associated with the induced interaction $-\left(\Delta N_\sigma\right)^2/\mathcal{N}$, such that one obtains
\begin{align} \label{eq:Fnex}
    F_{n,\sigma\sigma} - \new{f^{(0)}}\simeq\pm \frac{\left(\Delta N_\sigma\right)^2}{\mathcal{N}}\,,
\end{align}
with the $+$ and $-$ signs applying to fermionic and \textit{non-degenerate} ($\p_1\neq \p_2)$ bosonic impurities, respectively. 
The corollary of this is that the mediated interactions vanish between distinguishable impurities: $F_{n,\up\down} - \new{f^{(0)}}\simeq 0$. 
Taken together with Eq.~\eqref{eq:FmuFn}, we find that Eq.~\eqref{eq:Fnex} immediately implies that 
\begin{equation}
\label{eq:Fmunondeg}
    F_{\mu,\sigma\sigma'}\simeq 
    \begin{cases}
    0 & \text{fermions} \\ \new{f^{(0)}}-2 \frac{\left(\Delta N_\sigma\right)^2}{\mathcal{N}} & \text{bosons }(\p_1\neq\p_2) \\[0.5em] \new{f^{(0)}}-\frac{\Delta N_{\up}\Delta N_{\down}}{\mathcal{N}} & \text{distinguishable}
    \,.
    \end{cases}
\end{equation}
For fermionic impurities, the absence of interactions has already been observed in a harmonically trapped Fermi gas~\cite{Nascimbene2010} and explained theoretically in Ref.~\cite{Mora2010}. For the case of bosons, the factor-of-two statistical enhancement of the induced interaction in Eq.~\eqref{eq:Fmunondeg} was, to our knowledge, first proposed in Ref.~\cite{levinsen2025}, and is yet to be investigated experimentally. 

Equation~\eqref{eq:Fnex} can be shown to be exact in the perturbative limit of weak impurity-medium interactions~\cite{Mora2010,Yu2012}, as is also confirmed by our wave-function approach \new{below}. Indeed, in the case of fermionic impurities, this expression likely holds for arbitrary impurity-medium interaction strengths, since this is the only ($s$-wave) interaction between fermionic impurities at low energies. However, in general, Eq.~\eqref{eq:Fnex} only represents a subset of the possible processes that contribute to the polaron-polaron interactions. Most notably, it cannot capture the 
medium-induced interactions at fixed $n_\mathrm{med}$ when the bosonic impurities are \textit{degenerate} ($\p_1=\p_2$), where there is no concept of exchange. We emphasize that bosonic degeneracy is not an esoteric scenario: It can happen if the impurities form a condensate in thermodynamic equilibrium, or dynamically if they are injected at the same momenta.\footnote{\new{In practice, the uncertainty principle implies that impurities injected into a small volume $V\sim \Delta x^3$ should be treated as degenerate whenever $|\p_1-\p_2|\lesssim 1/2\Delta x$, such as is the case in multi-dimensional spectroscopy protocols in a ``box geometry"~\cite{Muir2022}.}} The latter can be via radiofrequency transfer into the interacting state from an atomic BEC or by converting photons from a laser into excitons in a semiconductor. In this case, the leading order effect of the medium is to enhance the (repulsive) bare interaction \new{$f^{(0)}$} between impurities, thus resulting in a repulsive rather than an attractive medium-induced interaction~\cite{levinsen2025}. On the other hand, if $\mu_{\rm med}$ rather than $n_{\rm med}$ is fixed, then we find that the subtracted term in Eq.~\eqref{eq:FmuFn} dominates at leading order and the overall interaction is attractive (see Table~\ref{tab:signs}). 

We also show here that a similar situation occurs for the case of two distinguishable impurities, where there is once again no exchange process. For fixed $\mu_{\rm med}$, the dominant behavior in the weak-coupling limit is captured by Eq.~\eqref{eq:Fmunondeg}, while for the case of fixed $n_{\rm med}$, the leading order effect of the medium is to modify or ``dress'' the bare interaction \new{$f^{(0)}$} between impurities. In this case, the influence of the medium depends on how each impurity interacts with the medium, as indicated in Table~\ref{tab:signs}, and it is independent of whether the impurity momenta are strictly identical. 

Finally, we stress that in Table~\ref{tab:signs} we have provided the signs of the \textit{dominant} medium-induced part of the polaron interactions at weak impurity-medium interactions. However, an effect such as medium-enhanced repulsion between impurities is very general, and will happen for all the cases where the impurities themselves interact, even if---in some cases---it is a subleading process. 
Similarly, there are other higher order processes that are not captured by Eq.~\eqref{eq:Fnex}. One such example is ``phase-space filling'' where impurities compete for dressing by majority particles, which appears to be a generic feature of polaron-polaron interactions in the semiconductor platform~\cite{Tan_PRX20,Muir2022}, while another is the formation of two-impurity bipolaron bound states due to the mediated interaction~\cite{Camacho_Guardian2018-prl,Naidon2018,Muir2022} which can even be of Efimovian character~\cite{Naidon2018,Enss2020}. On the other hand, the link between polaron interactions with the different medium constraints, Eq.~\eqref{eq:FmuFn}, is non-perturbative and completely general since it follows from thermodynamic arguments.

\subsection{Experimental protocols and measurements}

As illustrated in Table~\ref{tab:signs}, previous experiments on polaron-polaron interactions have effectively probed a range of different scenarios, which naturally complicates their direct comparison. In particular, one needs to carefully consider which constraints have been imposed on the medium response and how the impurities have been prepared, e.g., whether or not they are degenerate. Below we briefly discuss the different experiment protocols and highlight another key factor: that it is important to distinguish between measurements of equilibrium properties and those performed on fast timescales.

\subsubsection{``Injection'' of impurities in quantum gases}
A common class of cold-atom experiments on polarons involves using a radiofrequency pulse to transfer impurity atoms into a state that interacts with the medium, as illustrated in Fig.~\ref{fig:sketch2}(b). This so-called ``injection'' protocol probes the impurity energy spectrum and is typically carried out on short timescales such that the medium does not have time to reach chemical equilibrium. Hence, the impurities are only sensitive to their local environment (e.g., at the center of the harmonic trap), and the polaron-polaron interactions are effectively determined by the fixed local density. This explains why a recent experiment~\cite{Baroni2024} on thermal bosonic and fermionic impurities reported medium-induced interactions consistent with those expected for fixed $n_\mathrm{med}$ in the weak-coupling limit (Table~\ref{tab:signs}), even though the measurements were carried out in a harmonic trap where the local $\mu_\mathrm{med}$ at the trap center should be fixed by the rest of the gas in thermal equilibrium (see Fig.~\ref{fig:sketch2}). 

We emphasize that one can only access polaron-polaron interactions in the injection protocol by going beyond the linear response regime. Simply varying the initial density of impurities in the non-interacting state, while remaining within linear response, will not yield any polaron-polaron interactions, as was likely the case in Ref.~\cite{Scazza2017}. \new{Likewise, it is important to inject impurities sufficiently slowly that the system can reach a local thermodynamic equilibrium, otherwise dynamical effects such as retardation will play a crucial role. In other words, we require the impurity injection rate $\Gamma\lesssim c_s/l$ where $c_s$ is the speed of sound of the medium, and $l$ is the impurity separation.}

\subsubsection{Equilibrium quantum-gas experiments}
Cold-atom polaron experiments have also been carried out in equilibrium, where one instead measures thermodynamic quantities such as the equation of state from density profiles in the harmonic trap (Fig.~\ref{fig:sketch2})(a). In this case, we would expect that a description in terms of a fixed medium chemical potential would typically be appropriate. Indeed, equilibrium experiments carried out in harmonic traps have observed attractive mediated interactions between bosons in a Fermi gas~\cite{DeSalvo2019} and an absence of induced interactions for fermionic Fermi polarons~\cite{Nascimbene2010}, which is consistent with our results at fixed $\mu_\mathrm{med}$ in Table~\ref{tab:signs}. While the statistical enhancement for bosonic impurities in Eq.~\eqref{eq:FmuFn} has yet to be investigated, it could, for instance, be measured by comparing the results of degenerate and non-degenerate impurities in an equilibrium experiment similar to that of Refs.~\cite{DeSalvo2019,Cai2025}, or by comparing the results of equilibrium and out-of-equilibrium probes in a setup like in Ref.~\cite{Baroni2024}.

We note that it is possible to measure polaron-polaron interactions at fixed $n_\mathrm{med}$ in a trapped equilibrium gas if one monitors the local density like in Ref.~\cite{Schirotzek2009}. However, this may be difficult to achieve outside the unitary scale-invariant regime ($1/a = 0$) considered in Ref.~\cite{Schirotzek2009}, since variations in the local density $n$ within the trap also generally modify the local interaction parameter $n a^3$, which cannot be scaled away.

\new{A natural question is what timescale $t$ is required to reach thermal equilibrium, given the dilute impurity concentration. Similarly to above, a simple estimate requires
an equilibration time $t \gtrsim l/c_s$.}

\subsubsection{Optically introduced polarons in 2D semiconductors}

In the semiconductor platform, impurities such as excitons or exciton polaritons are introduced optically~\cite{Tan_PRX20,Muir2022,Tan_PRX2023,Ni2025} and thus the measurement of polaron-polaron interactions is similar to the injection protocol in atomic gases. Therefore, even in the cases where the background medium has a chemical potential (e.g., due to a gate voltage or a pump laser), the measurement is sufficiently fast that the medium has no time to reach any thermal equilibrium and it should thus be treated as having a local fixed $n_\mathrm{med}$.  

The impurities in recent experiments are either distinguishable or identical degenerate excitons~\cite{Muir2022,Ni2025} or indistinguishable exciton-polaritons~\cite{Tan_PRX20,Tan_PRX2023}, and according to Table~\ref{tab:signs} we therefore expect repulsive induced polaron interactions.\footnote{Note that the direct interaction $g_{\sigma\sigma}$ is slightly modified in the 2D case since it will depend logarithmically on the exciton density or the exciton-polariton energy, in addition to the exciton-exciton scattering length~\cite{Levinsen2Dreview}.} This is consistent with the observations for the case of attractive polarons~\cite{Tan_PRX20,Muir2022,Tan_PRX2023}. On the other hand, Ref.~\cite{Tan_PRX2023} reported attractive interactions for the metastable repulsive branch, a result which was interpreted in terms of a phase-space filling effect. While we do not find such a mechanism up to fourth order in the impurity-medium interaction strength~\cite{levinsen2025}, it could potentially be a higher-order effect that is beyond the scope of the present work.

\new{\subsection{Quasiparticle interactions versus effective interaction potentials
\label{sec:QPvsInd}}
Finally, we want to make a clear distinction between the quasiparticle interaction, as defined in Landau's Fermi liquid theory~\cite{BaymPethick1991book}, and an effective mediated interaction potential between impurities~\cite{ando2025}. The key point is that the quasiparticle interaction $f_{\p_1\sigma,\p_2\sigma'}$ in Eq.~\eqref{eq:energydensity1} defines the \textit{energy shift} of one quasiparticle due to the presence of another. Thus, the impurity momenta $\p_1$, $\p_2$ are unchanged in this interaction and the impurities can, at most, exchange their momenta if they are indistinguishable ($\sigma = \sigma'$). This also implicitly assumes that the quasiparticles remain well defined and are adiabatically connected to bare impurities.

By contrast, a mediated interaction potential between impurities in a uniform system only needs to conserve the total momentum $\p_1 + \p_2$, not the impurity momenta separately, and is defined independently of the impurity statistics. Such a potential $V_{\rm med}(\q,\omega)$ depends on the momentum $\q$ and energy $\omega$ transferred between impurities (and also on the momenta $\p_1$, $\p_2$ in general). While this potential can in principle be used to calculate the quasiparticle interaction, it must be converted into a scattering amplitude in order to yield an interaction energy shift (similarly to how atomic potentials are converted into scattering lengths in order to obtain mean-field interaction shifts). Furthermore, an effective mediated interaction can give rise to bound states between impurities, which go beyond the simple quasiparticle picture envisioned by Eq.~\eqref{eq:energydensity1}.

A possible source of confusion is that there are limits where the quasiparticle interaction recovers the mediated interaction potential in the static limit $\omega \to  0$ (see Sections \ref{sec:bosepolaron} and \ref{sec:fermipolaron}). This is because of two factors. First, in the  limit of weak impurity-medium interactions, the interaction energy shift due to the medium reduces to the leading-order term in the Born series, which is directly connected to the weak-coupling interaction potential, i.e., for non-degenerate non-interacting bosonic or fermionic impurities, $f_{\p_1\sigma,\p_2\sigma} = \pm V_{\rm med}(\p_1 - \p_2,\epsilon_{\p_1\sigma}-\epsilon_{\p_2\sigma})$, respectively~\cite{Paredes2024}, with the direct term $V_{\rm med}(\0,0)$ vanishing at fixed medium density. Second, in the limit of infinite impurity mass, where the impurity kinetic energy $\epsilon_{\p\sigma} \to 0$, the constraints on impurity momenta are relaxed due to the absence of recoil, such that the interactions only depend on the momentum transfer between impurities, i.e., the  potential $V_{\rm med}(\p_1 - \p_2,0)$. Thus, the quasiparticle interaction becomes insensitive to any momentum conservation and can be directly related to the static potential in these limits. Indeed, many formulations of mediated interactions are between fixed impurities, including the original RKKY interaction~\cite{Giuliani-Vignale-book2008}. We remark that such a static potential can also be achieved for finite-mass impurities, but only in the special case where $\epsilon_{\p_1\sigma}=\epsilon_{\p_2\sigma}$, corresponding to momenta with the same magnitude.}

\section{Constraints on the medium response}
\label{sec:ensemble}
As discussed above and illustrated in Fig.~\ref{fig:sketch}, the effective constraints on how the medium can respond to a perturbation have profound effects on the interactions between polarons. For instance, in liquid Helium mixtures the polaron interactions are typically extracted from equilibrium thermodynamic quantities at fixed pressure, and theories therefore model the system at a fixed medium chemical potential~\cite{Bardeen1966,Bardeen1967,Saam1969}. On the other hand, in the cold-atom context either type of medium constraint can apply: While the medium chemical potential is effectively held fixed in the case of equilibrium experiments in trapped geometries where the local density approximation holds, experiments on short timescales (such as those where impurities are injected using radiofrequency spectroscopy) can effectively probe the medium at a fixed density. The dynamical injection of impurities in cold atomic gases is in turn closely related to the optical probes used in the semiconductor context, where absorption, reflection, and transmission spectroscopies are all effectively conducted at a fixed medium density. These considerations motivate a general discussion of how the polaron interactions with different medium constraints are related. 

\subsection{Bosonic impurities}
\label{sec:ensemblea}

Let us for simplicity first consider a quantum mixture with a single minority bosonic component denoted $\sigma$. The expansion of the energy density in powers of the impurity density \new{now} takes the general form:
\begin{align}\label{eq:energycanonical}
    \mathcal{E}(n,n_{\sigma})=\mathcal{E}_0+E_{\mathrm{pol},\sigma}n_{\sigma}+\frac12F_{n,\sigma\sigma} n_{\sigma}^2
    \,,
\end{align}
where, \new{as above,} $\mathcal{E}_0$ is the energy density of the medium in the absence of impurities, and $E_{\mathrm{pol},\sigma}$ and $F_{n,\sigma\sigma}$ are the \new{momentum-independent} polaron energy and polaron-polaron interaction constant, respectively, \new{see the discussion in Section~\ref{sec:summary}}. To simplify the notation, we henceforth denote the medium density by $n=n_\mathrm{med}$. 

Importantly, the polaron energy and the polaron-polaron interaction constant both depend on $n$ but not on $n_{\sigma}$. That is, we have 
\begin{subequations}
\begin{align}
    E_{\mathrm{pol},\sigma} & = \left.\pdv{\mathcal{E}}{n_{\sigma}}\right|_{n_{\sigma}\to0}\,, \\
    F_{n,\sigma\sigma}&=\left.\pdv{^2\mathcal{E}}{n_{\sigma}^2}\right|_{n_{\sigma}\to0}\,.
\end{align}
\end{subequations}
We note that Eq.~\eqref{eq:energycanonical} is an exact expansion in the limit $n_{\sigma} \to 0$ and that it can even describe systems with bipolarons (bound states of two impurities due to the interaction with the medium) in which case the bipolaron binding energy should be incorporated into $E_{\mathrm{pol},\sigma}$.

In a scenario where the medium chemical potential is kept fixed rather than the density, we instead consider a mixed ensemble, where we have the grand canonical potential for the medium:
\begin{align}\label{eq:OmegafromE}
    \Omega(\mu,n_{\sigma})=\mathcal{E}(n,n_{\sigma})-n\mu\,. 
\end{align}
For simplicity, we write the medium chemical potential as $\mu = \mu_\mathrm{med}$.  Taking the derivative with respect to $n_\sigma$ yields
\begin{align}\label{eq:Epolensemble}
    \pdv{\Omega}{n_{\sigma}}
    & =
    \pdv{\mathcal{E}}{n_{\sigma}}\,,
\end{align}
where in the last step we used $\frac{\partial \Omega}{\partial n} = \frac{\partial \mathcal{E}}{\partial n} -\mu = 0$ which follows from the definition of the medium chemical potential, $\mu=\pdv{\mathcal{E}}{n}$. Importantly, this implies that the polaron energy is the same within the two ensembles, as expected.

To obtain the quasiparticle interactions at fixed $\mu$, $\left.F_{\mu,\sigma\sigma}=\pdv{^2\Omega}{n_\sigma^2}\right|_{n_\sigma\to0}$, we take the second derivative which yields
\begin{align}\label{eq:Omega2nd}
    \pdv{^2\Omega}{n_\sigma^2} =\pdv{^2\mathcal{E}}{n_\sigma^2}+\pdv{^2\mathcal{E}}{n\,\partial n_\sigma}\pdv{n}{n_\sigma}\,,
\end{align}
Furthermore, in order to ensure that the medium chemical potential stays fixed when we add a small amount of impurities we must have~\cite{Massignan2005}
\begin{align}\label{eq:deltamu}
    \delta \mu=\left(\pdv{^2\mathcal{E}}{n\partial n_{\sigma}}+\pdv{^2\mathcal{E}}{n^2}\pdv{n}{n_{\sigma}}\right)\delta n_{\sigma}=0\,.
\end{align}
Thus we finally obtain~\cite{levinsen2025} 
\begin{align}\label{eq:FmuFn1}
    F_{\mu,\sigma\sigma} & = F_{n,\sigma\sigma}-\pdv{^2\mathcal{E}}{n^2}\left(\pdv{n}{n_{\sigma}}\right)^2=F_{n,\sigma\sigma} - \frac{(\Delta N_\sigma)^2}{\mathcal{N}}\,,
\end{align}
where we evaluate the derivatives at $n_\sigma\to0$. Here we have introduced the density of states at the chemical potential
\begin{align}\label{eq:N}
\mathcal{N}=\pdv{n}{\mu}\,, 
\end{align}
as well as the number of particles in the impurity dressing cloud
\begin{align}\label{eq:DeltaN}
    \Delta N_\sigma=\pdv{n}{n_{\sigma}}=- \pdv{E_{\mathrm{pol},\sigma}}{\mu}\,.
\end{align}
The last step follows from $n=-\pdv{\Omega}{\mu}$ and $\mu_\sigma=\pdv{\Omega}{n_\sigma}$, with $\mu_\sigma=E_{\mathrm{pol},\sigma}$ in the limit $n_\sigma\to0$.

We emphasize that the above arguments are based purely on thermodynamic considerations. They are therefore independent of the medium statistics, dimensionality, and specific physical realization of a highly imbalanced mixture. We now show that these arguments also carry over straightforwardly to fermionic and distinguishable impurities.

\subsection{Fermionic impurities}
\label{sec:ensemblea2}

The main complication in the case of fermionic impurities is that, even in the absence of interactions, these have a kinetic energy at zero temperature. In other words, for a single minority fermionic component $\sigma$, the energy density takes the Landau-Pomeranchuk form
\begin{align}\label{eq:energycanonicalFermi}
    \mathcal{E}(n,n_{\sigma})=\mathcal{E}_0+E_{\mathrm{pol},\sigma}n_{\sigma}+\frac12F_{n,\sigma\sigma} n_{\sigma}^2+\mathcal{E}_{\mathrm{kin},\sigma}\,,
\end{align}
where $\mathcal{E}_{\mathrm{kin},\sigma} = \frac{3}{5} E_{F,\sigma} n_{\sigma} = \frac35 \frac{(6\pi^2)^{2/3}}{2m^*_\sigma}n_{\sigma}^{5/3}$, with $m^*_\sigma$ the polaron effective mass and $E_{F,\sigma}$ the corresponding Fermi energy. 

The argument proceeds nearly as above. The quasiparticle interactions are now
\begin{subequations}
\begin{align}
    F_{n,\sigma\sigma}&=\left.\pdv{^2\mathcal{E}'}{n_{\sigma}^2}\right|_{n_{\sigma}\to0}\,, \\
    F_{\mu,\sigma\sigma}&=\left.\pdv{^2\Omega'}{n_{\sigma}^2}\right|_{n_{\sigma}\to0}\,,
\end{align}
\end{subequations}
where we \new{use} $\mathcal{E}'=\mathcal{E}-\mathcal{E}_{\mathrm{kin},\sigma}$ and $\Omega'=\Omega-\mathcal{E}_{\mathrm{kin},\sigma}$ since we require the $O(n_{\sigma}^2)$ term. Now we have
\begin{align}
    \pdv{^2\Omega'}{n_{\sigma}^2}=\pdv{^2\mathcal{E}'}{n_{\sigma}^2}+\pdv{^2\mathcal{E}}{n\,\partial n_\sigma}\pdv{n}{n_\sigma}-\pdv{^2\mathcal{E}_{\mathrm{kin},\sigma}}{n\,\partial n_\sigma}\pdv{n}{n_\sigma}\,,
\end{align}
where we have again used Eq.~\eqref{eq:Epolensemble}. Importantly, the last term vanishes when we take $n_\sigma$ to zero, and therefore the presence of the impurity kinetic energy does not materially change the relationship compared with \eqref{eq:Omega2nd}. As above, making use of Eq.~\eqref{eq:deltamu} we find the same relationship between the quasiparticle interactions, namely
\begin{align}
    F_{\mu,\sigma\sigma} & =F_{n,\sigma\sigma} - \frac{(\Delta N_\sigma)^2}{\mathcal{N}}\,.
\end{align}

\subsection{Multiple impurity components}
\label{sec:ensembleb}

The above arguments are straightforward to extend to multiple impurity components. To be specific, we consider a three-species mixture with $\sigma =\{\up,\down\}$ and $n_{\up},n_{\down}\ll n$. This could, for instance, be a dilute solution of the two spin components of $^3$He in $^4$He, for which the corresponding formalism was developed in Ref.~\cite{Bardeen1967}. The energy density is then of the form
\begin{align}
    \mathcal{E}(n,n_{\up},n_{\down})=&\mathcal{E}_0+E_\mathrm{pol,\up}n_{\up}+E_\mathrm{pol,\down}n_{\down}
    +F_{n,\up\down} n_{\up}n_{\down} + \ldots
    \,,
\end{align}
where the medium energy density $\mathcal{E}_0$, the two polaron energies $E_{\mathrm{pol},\up}$ and $E_{\mathrm{pol},\down}$, and the polaron interaction constant $F_{n,\up\down}$ all depend on the medium density $n$ but not on $n_{\up}$ and $ n_{\down}$. Here we have dropped the higher powers of $n_\up$ and $n_{\down}$ since we are focused on the quasiparticle interactions between distinguishable impurities, and the next-order contributions simply correspond to the interaction and/or kinetic energy terms considered in the previous sections. 

We once again introduce the grand canonical potential for the medium
\begin{align}
    \Omega(\mu,n_{\up},n_{\down})=\mathcal{E}(n,n_{\up},n_{\down})-\mu n\,,
\end{align}
from which we can define the polaron interaction at fixed medium chemical potential:
\begin{align}
    F_{\mu, \up\down}=\left.\pdv{^2\Omega}{n_\up  \partial n_{\down}}\right|_{n_\up ,n_{\down} \to0}\,.
\end{align}
Going through the same steps as above, we find
\begin{align}\label{eq:FmuFnij}
    F_{\mu,\up\down}
    & =F_{n,\up\down} + \pdv{E_{\mathrm{pol},\up}}{n}\pdv{n}{n_{\down}} =F_{n,\up\down} + \pdv{E_{\mathrm{pol},\down}}{n}\pdv{n}{n_\up}\,.
\end{align}

The final argument is also the same. In order to ensure that the medium chemical potential stays fixed upon adding a small amount of $\up$ and $\down$ impurities, we require
\begin{align}
    \delta \mu=\!\left(\pdv{^2\mathcal{E}}{n\partial n_{\up}}\!+\!\pdv{^2\mathcal{E}}{n^2}\pdv{n}{n_{\up}}\right)\!\delta n_{\up}\!+\!\left(\pdv{^2\mathcal{E}}{n\partial n_{\down}}\!+\!\pdv{^2\mathcal{E}}{n^2}\pdv{n}{n_{\down}}\right)\!\delta n_{\down}=0\,.
\end{align}
Therefore, using the definition of the impurity chemical potentials, $\mu_{\sigma}=\pdv{\mathcal{E}}{n_{\sigma}}$, and imposing that the terms in the brackets vanish, we have
\begin{align}
    \pdv{\mu_{\sigma}}{n}=-\pdv{\mu}{n}\pdv{n}{n_{\sigma}} \,.
\end{align}
Once again, recognizing that $\mu_{\sigma}=E_\mathrm{pol,\sigma}$ when $n_{\sigma} \to0$, we finally obtain
\begin{align}
    F_{\mu,\up\down}& =F_{n,\up\down}- \pdv{\mu}{n}\left(\pdv{E_{\mathrm{pol},\up}}{\mu}\right)\left(\pdv{E_{\mathrm{pol},\down}}{\mu}\right)\,.
\end{align}
This can also be written as
\begin{align}\label{eq:FmuFn2}
    F_{\mu,\up\down}& =F_{n,\up\down} - \frac{\Delta N_\up  \Delta N_{\down}}{\mathcal{N}}\,,
\end{align}
with the number of particles in the dressing cloud $\Delta N_{\sigma}$ defined in Eq.~\eqref{eq:DeltaN}.

Equation~\eqref{eq:FmuFn2} illustrates how the effective polaron interaction between two distinguishable impurities at fixed medium chemical potential contains a contribution from the combined effect of the two dressing clouds. This will lead to an attraction in the cases where both impurities attract or repel medium particles. However, it can also lead to an effective repulsion if the impurity particles interact with the medium with opposite signs, i.e., if one impurity forms an attractive polaron while the other forms a repulsive polaron. In other words, the change in medium density due to the presence of one impurity can raise the energy of the other impurity if they have opposing dressing clouds. 

\subsection{Comparison with theories of liquid He mixtures}
\label{sec:helium}

To wrap up our discussion of the different constraints on the medium response, we find it useful to refer back to the literature on liquid He mixtures, specifically the case of a dilute admixture of $^3$He in $^4$He~\cite{Bardeen1966,Bardeen1967,Saam1969}. Here, experiments are conducted at fixed pressure, which is equivalent to working at a constant medium chemical potential in a fixed volume since the pressure $P=-\Omega(\mu)/V$.

Reference~\cite{Bardeen1967} distinguishes between two separate contributions to the quasiparticle interactions. The ``direct'' interaction between two distinguishable $^3$He atoms ($\up$ and $\down$) is
\begin{align}
    V_\mathrm{dir}&=\pdv{\mu_{3\up}}{n_{3\down}},
\label{eq:He-dir}
\end{align}
where we use the subscripts to indicate a particular spin-component of $^3$He (and similarly to indicate $^4$He below). Comparing with our calculations for two distinguishable impurity components in Section~\ref{sec:ensembleb} above, where we have $F_{n,\up\down}=\pdv{^2\mathcal{E}}{n_{\up}\partial n_{\down}}=\pdv{\mu_{\up}}{n_{\down}}=\pdv{\mu_{\down}}{n_{\up}}$ (evaluated in the limit $n_{\up},n_{\down}\to0$), we see that the direct contribution precisely corresponds to the quasiparticle interaction at fixed medium density.

The second contribution is the so-called ``phonon-induced'' part of the interaction, which is
\begin{align}
    V_\mathrm{ind}&=\pdv{\mu_{3\up}}{n_4} \pdv{n_4}{n_{3\down}}.
\label{eq:He-ind}
\end{align}
Note that these derivatives are carried out at fixed pressure, and therefore at fixed $\mu_4$. In our notation, the ``induced interaction'' corresponds to $\pdv{\mu_{\up}}{n} \pdv{n}{n_{\down}}=\pdv{\mu_{\down}}{n} \pdv{n}{n_{\up}}$ which, according to Eq.~\eqref{eq:FmuFnij}, is precisely the additional contribution that arises at fixed medium chemical potential.

All in all, we find that the quasiparticle interaction between $\up$ and $\down$ $^3$He impurities is
\begin{align}
    F_{\mu,\up\down}=V_\mathrm{dir}+V_\mathrm{ind}.
\end{align}
This makes it clear that theories that calculate the mediated quasiparticle interactions at fixed medium density---such as in Refs.~\cite{YuPethick2012,Camacho2018} for the case of indistinguishable impurities---are, in effect, calculating a different quantity from the so-called induced interactions introduced in the context of He mixtures~\cite{Bardeen1967}.

\section{Impurities in a Bose gas}
\label{sec:bosepolaron}

We now turn to the explicit evaluation of polaron interactions, tackling first the case of the Bose polaron where the medium corresponds to a weakly interacting BEC at $T=0$. The single-Bose-polaron case has previously been investigated using a multitude of different theoretical techniques such as the Fr\"ohlich model~\cite{Devreese2009,GrusdtDemler2015}, perturbation theory~\cite{Novikov2009,Huang2009,Casteels2014,Christensen2015}, variational and diagrammatic methods~\cite{Rath2013,Li2014,Levinsen2015}, Monte-Carlo simulations~\cite{Ardila2015,Ardila2016}, coherent-state approaches~\cite{Shchadilova2016}, and Gross-Pitaevskii-based theories~\cite{Massignan2021}. Here we consider the limit of weak impurity-medium interactions such that we can perform a rigorous perturbative expansion in the impurity-medium interaction strengths up to second order.

Describing the medium within Bogoliubov theory and considering an arbitrary number of impurity species of arbitrary statistics, we then have the Hamiltonian
\begin{multline}\label{eq:HamFrohlich}
    \hat{H} =
    \sum_{\k\neq0} E_\k \beta^\dag_\k \beta_\k + \sum_{\k,\sigma}(\epsilon_{\k \sigma}+ g_{b\sigma} n) c_{\k \sigma}^\dag c_{\k \sigma}^{} + \sum_\sigma \frac{g_{b\sigma} \sqrt{N}}{V} \sum_{\k\neq0,\p}W_\k \, c^\dag_{\p+\k \sigma} c_{\p \sigma}^{} \left(\beta_\k^{} + \beta^\dag_{-\k} \right) \\ 
    + \sum_{\sigma\sigma'}\frac{g_{\sigma\sigma'}}{2V} \sum_{\k\k'\q} c^\dag_{\k \sigma} c^\dag_{\k'\sigma'} c_{\k'+\q \sigma'}^{} c_{\k-\q \sigma}^{} \, ,
\end{multline}    
which is taken to be relative to the energy of the medium Bose gas with particle number $N$ in a volume $V$ (giving the density $n = N/V$). In the limit of a single impurity, Eq.~\eqref{eq:HamFrohlich} reduces to the well-known Fr{\"o}hlich model~\cite{GrusdtDemler2015}. However, the last term allows us to include the effect of bare impurity interactions, which exist in the case of bosonic or distinguishable impurities.

In Eq.~\eqref{eq:HamFrohlich}, the operators $\beta_\k$ and $\beta^\dag_\k$ are the usual Bogoliubov operators for the medium particles with dispersion $E_\k = \sqrt{\epsilon_{\k b}(\epsilon_{\k b} + 2 g_{bb} n)}$, where $\epsilon_{\k b} = |\k|^2/2{m_b} \equiv k^2/2m_b$ with $m_b$ the boson mass. Likewise, $c^\dag_{\k \sigma}$ and $c_{\k \sigma}^{}$ respectively create and destroy impurities of species $\sigma$ at momentum $\k$ with mass $m_\sigma $ and dispersion $\epsilon_{\k \sigma} = k^2/2m_\sigma $. We have also introduced the impurity-impurity, boson-impurity, and boson-boson interaction coefficients that, respectively, take the forms $g_{\sigma\sigma'}=2\pi a_{\sigma\sigma'}/{m_{\sigma\sigma'}}$, $g_{b\sigma}=2\pi a_{b\sigma}/{m_{b\sigma}}$, and $g_{bb}=4\pi a_{bb}/{m_b}$. These are written in terms of the corresponding scattering lengths $a_{\sigma\sigma'}$, $a_{b\sigma}$, and $a_{bb}$, and we have introduced the reduced impurity-impurity mass $m_{\sigma\sigma'}=m_\sigma m_{\sigma'}/(m_\sigma +m_{\sigma'})$ and impurity-boson mass $m_{b\sigma}=m_bm_\sigma /(m_b+m_\sigma )$. We work at $T=0$ and assume the medium to be dilute, i.e., $na_{bb}^3\ll1$, with $a_{bb}>0$ for stability. Writing the impurity-boson interaction in terms of Bogoliubov operators gives the additional function $W_\k = \sqrt{\epsilon_{\k b}/E_\k}$. 

We emphasize that our results below can be straightforwardly generalized to other scenarios featuring Bose polarons. For instance, with minimal modifications, our theory of polaron interactions can be extended to describe impurities in 2D ultracold atomic gases~\cite{Ardila2020,CardenasCastillo2023,Nakano2024} or excitonic (and potentially electronic) impurities in an atomically thin semiconductor with either an exciton or an exciton-polariton reservoir~\cite{Takemura_NaturePhys2014,Levinsen_PRL2019,Tan_PRX2023,Choo2024}.

\subsection{Single-polaron problem}
\label{sec:single-Bose-polaron}
For completeness, we first consider a single impurity of species $\sigma$ and momentum $\p$ in a BEC. To describe this Bose polaron, we use the variational ansatz introduced in Ref.~\cite{Li2014} (see also Refs.~\cite{Rath2013,Levinsen2015}):
\begin{align}
    \ket{\Psi}=\bigg(\alpha_{\p}c^\dag_{\p \sigma}+\sum_{\k\neq0}\alpha_{\k\p}c^\dag_{\p+\k \sigma}\beta^\dag_{-\k}\bigg)\ket{\Phi}\,,
\end{align}
where $\ket{\Phi}$ is the medium BEC at density $n$. This describes how the impurity can create Bogoliubov excitations in the BEC, with such processes truncated at the level of a single excitation. The corresponding equations of motion are obtained by taking $\partial_{\lambda^*}\expval{(E-\hat H)}{\Psi}=0$, where $\lambda$ is any of the variational parameters $\alpha_{\p}$ and $\alpha_{\k\p}$. This procedure yields
\begin{subequations}
\begin{align}
    (E-\epsilon_{\p \sigma}-g_{b\sigma}n)\alpha_{\p} &= \frac{g_{b\sigma}\sqrt{N}}V\sum_{\k\neq0} W_\k \alpha_{\k\p}\,,\label{eq:eomalphasingle}\\
    (E-\epsilon_{\p+\k \sigma}-E_\k-g_{b\sigma}n)\alpha_{\k\p}&=\frac{g_{b\sigma}\sqrt{N}}{V}W_\k \alpha_{\p}\,.\label{eq:eomrhosingle}
\end{align}
\end{subequations}

To obtain the polaron energy to second order, we insert Eq.~\eqref{eq:eomrhosingle} in \eqref{eq:eomalphasingle} to find
\begin{align}
    E=\epsilon_{\p \sigma}+g_{b\sigma}n+\frac{g_{b\sigma}^2n}{V}\sum_{\k\neq0}\frac{W_\k^2}{E-g_{b\sigma}n-\epsilon_{\p+\k \sigma}-E_\k}\,.
\end{align}
At the level of perturbation theory, we can \new{use the leading-order quasiparticle dispersion} $E\simeq g_{b\sigma}n+\epsilon_{\p \sigma}$ in the iterated term, \new{since corrections to the polaron energy and effective mass are higher order in $g_{b\sigma}$.} Thus, we find the polaron energy at momentum $\p$
\begin{align}\label{eq:EpolBose}
    E_{\mathrm{pol},\sigma}(\p)&\!=\!\epsilon_{\p \sigma}\!+\!g_{b\sigma}n\!+\!\frac{g_{b\sigma}^2n}V\sum_{\k\neq0}\frac{W_\k^2}{\epsilon_{\p \sigma}-\epsilon_{\p+\k \sigma}-E_\k}\,.
\end{align}
Finally, since the sum on $\k$ is divergent we follow the usual renormalization procedure by replacing $g_{b\sigma}\mapsto g_{b\sigma}+(g_{b\sigma}^2/V)\sum_{\k}1/\bar\epsilon_{\k \sigma}$ at lowest order (we define $\bar\epsilon_{\k \sigma}=\epsilon_{\k \sigma}+\epsilon_{\k b}$), which gives
\begin{align}\label{eq:EpolBose2}
    E_{\mathrm{pol},\sigma}(\p)&\!=\!\epsilon_{\p \sigma}\!+\!g_{b\sigma}n\!+\!\frac{g_{b\sigma}^2n}V\sum_{\k\neq0}\left(\frac{W_\k^2}{\epsilon_{\p \sigma}-\epsilon_{\p+\k \sigma}-E_\k}\!+\!\frac1{\bar\epsilon_{\k \sigma}}\right)\,.
\end{align}
This expression precisely matches the polaron energy at second order evaluated in previous works~\cite{Novikov2009,Casteels2014,Christensen2015}, and it can also be used to find the leading-order correction to the polaron effective mass. 

\subsection{Polaron interactions at fixed medium density}
We now investigate the general scenario of two arbitrary impurities immersed in a weakly interacting Bose gas. For clarity, we will first neglect any direct interactions between the impurities, in which case we will find that exchange is the dominant polaron interaction (as long as such processes exist). We then introduce an impurity-impurity short-range interaction, which leads to a medium-modified interaction that turns out to be the dominant effect for those cases where there is no exchange~\cite{levinsen2025}.

To be specific, we will consider two impurities of species $\sigma$ and $\sigma'$ with momenta $\p_1$ and $\p_2$, respectively, where both the species and the momenta can be the same or distinct. In general, the \new{two-quasiparticle energy takes} the form
\begin{align}\label{eq:Fnp1p2}
    E(\p_1,\p_2)=E_{\mathrm{pol}, \sigma}(\p_1)+E_{\mathrm{pol}, \sigma'}(\p_2)+\frac{\new{f_{\p_1\sigma,\p_2\sigma'}}}V\,,
\end{align}
measured from that of the medium in the absence of the impurities. The function \new{$f_{\p_1\sigma,\p_2\sigma'}$ is the quasiparticle interaction introduced in Section~\ref{sec:summary}, and in the low-momentum limit it reduces to the quasiparticle interaction constant $F_{n,\sigma\sigma'}$ discussed in Sections ~\ref{sec:summary} and \ref{sec:ensemble}.}

\subsubsection{Non-interacting impurities}
We first consider the case where $g_{\sigma\sigma'}=0$, i.e., any effective interaction between the impurities will be purely mediated by the medium. To describe how the two impurities interact via the exchange of medium excitations, we generalize the two-impurity ansatz introduced in Refs.~\cite{Naidon2018,levinsen2025} to arbitrary momenta and impurity species, respectively. Within this ansatz, the impurities exchange a single Bogoliubov mode, leading to
\begin{multline}
\label{eq:state-ansatzsigma}
    \ket{\Psi}=\bigg[\alpha_{\p_1\p_2}c^\dag_{\p_1 \sigma}c^\dag_{\p_2\sigma'} +\sum_{\k\neq0}\rho_{\p_1\p_2\k}c^\dag_{\p_1 \sigma}\beta^\dag_{-\k}c^\dag_{\p_2+\k \sigma'} \\ 
    +(1-\delta_{\p_1\p_2}\delta_{\sigma\sigma'})\sum_{\k\neq0}\eta_{\p_1\p_2\k}c^\dag _{\p_1+\k \sigma}\beta^\dag_{-\k}c^\dag_{\p_2\sigma'}\bigg]\ket{\Phi}\,.
\end{multline} 
Note that the two terms featuring a Bogoliubov excitation are identical in the particular case where the two impurities are indistinguishable particles with exactly the same momenta. Therefore, we explicitly exclude this scenario from the ansatz. Furthermore, we emphasize that if the two impurities are indistinguishable fermions then the condition $\p_1=\p_2$ cannot be realized due to the Pauli principle, and therefore this case simply does not exist. 

The normalization of the state $\ket{\Psi}$ is needed in order to evaluate the equations of motion. Explicitly, the normalization requires
\begin{multline}
\label{eq:norm2imp}
    1=\braket{\Psi}{\Psi} =(1+\delta_{\sigma\sigma'}\delta_{\p_1\p_2})|\alpha_{\p_1\p_2}|^2 
    +\sum_{\k\neq0}|\rho_{\p_1\p_2\k}|^2 + (1-\delta_{\sigma\sigma'}\delta_{\p_1\p_2})\sum_{\k\neq0}|\eta_{\p_1\p_2\k}|^2 \\ 
    \pm \delta_{\sigma\sigma'} (1-\delta_{\p_1\p_2})
    \left(|\rho_{\p_1\p_2,\p_1-\p_2}|^2 + |\eta_{\p_1\p_2,\p_2-\p_1}|^2\right)\,, 
\end{multline}
where the terms with $\pm$ account for the Bose (upper sign) or Fermi (lower sign) statistics of identical impurities when $\sigma=\sigma'$.

As in the single-polaron case, taking $\partial_{\lambda*}\bra{\Psi} (E-\hat{H}) \ket{\Psi}=0$ with $\lambda\in\{ \alpha_{\p_1\p_2},\rho_{\p_1\p_2\k},\eta_{\p_1\p_2\k}\}$, we find the equations of motion
\begin{subequations}
\begin{align}
    E_{\p_1 \sigma;\p_2 \sigma'}\alpha_{\p_1\p_2} & =  \frac{g_{b  \sigma'}\sqrt{N}}V\sum_{\k\neq0}   W_{\k} \rho_{\p_1\p_2\k}
  +  (1-\delta_{\sigma\sigma'}\delta_{\p_1\p_2})\frac{g_{b  \sigma}\sqrt{N}}V\sum_{\k\neq0} W_{\k} \eta_{\p_1\p_2\k}  \nonumber \\ &  \hspace{-20mm}\pm \delta_{ \sigma\sigma'}(1-\delta_{\p_1\p_2})\left(\frac{g_{b  \sigma'}\sqrt{N}}VW_{\p_1-\p_2}\rho_{\p_1\p_2,\p_1-\p_2}
    +\frac{g_{b  \sigma}\sqrt{N}}VW_{\p_2-\p_1}\eta_{\p_1\p_2,\p_2-\p_1}\right),\label{eq:eomalphasigma}
\\
    E_{\p_1 \sigma;\p_2+\k \sigma';\k}\rho_{\p_1\p_2\k}&=  (1+\delta_{\sigma\sigma'}\delta_{\p_1\p_2}) \frac{g_{b  \sigma'}\sqrt{N}}{V}W_{\k} \alpha_{\p_1\p_2},\label{eq:eomrhosigma}\\
    E_{\p_1+\k \sigma;\p_2 \sigma';\k}\eta_{\p_1\p_2\k}&= \frac{g_{b  \sigma}\sqrt{N}}{V}W_{\k} \alpha_{\p_1\p_2}.\label{eq:eometasigma}
\end{align}
\end{subequations}
Here, we have used the notation $E_{\p_1\sigma;\p_2\sigma'}=E-\epsilon_{\p_1\sigma}-\epsilon_{\p_2\sigma'}-g_{b\sigma}n-g_{b\sigma'}n$, and $E_{\p_1\sigma;\p_2\sigma';\k}=E-\epsilon_{\p_1\sigma}-\epsilon_{\p_2\sigma'}-E_{\k} -g_{b\sigma}n-g_{b\sigma'}n$. We once again stress that the terms $\eta_{\p_1\p_2\k}$ are absent from the ansatz \eqref{eq:state-ansatzsigma} for identical impurities at equal momenta $\p_1=\p_2$, and thus the last equation \eqref{eq:eometasigma} also disappears since it originates from the minimization condition $\partial_{\eta^*_{\p_1\p_2\k}}\bra{\Psi} (E-\hat{H}) \ket{\Psi}=0$. 

\begin{figure}[tbp]    
\centering
\includegraphics[width=.65\linewidth]{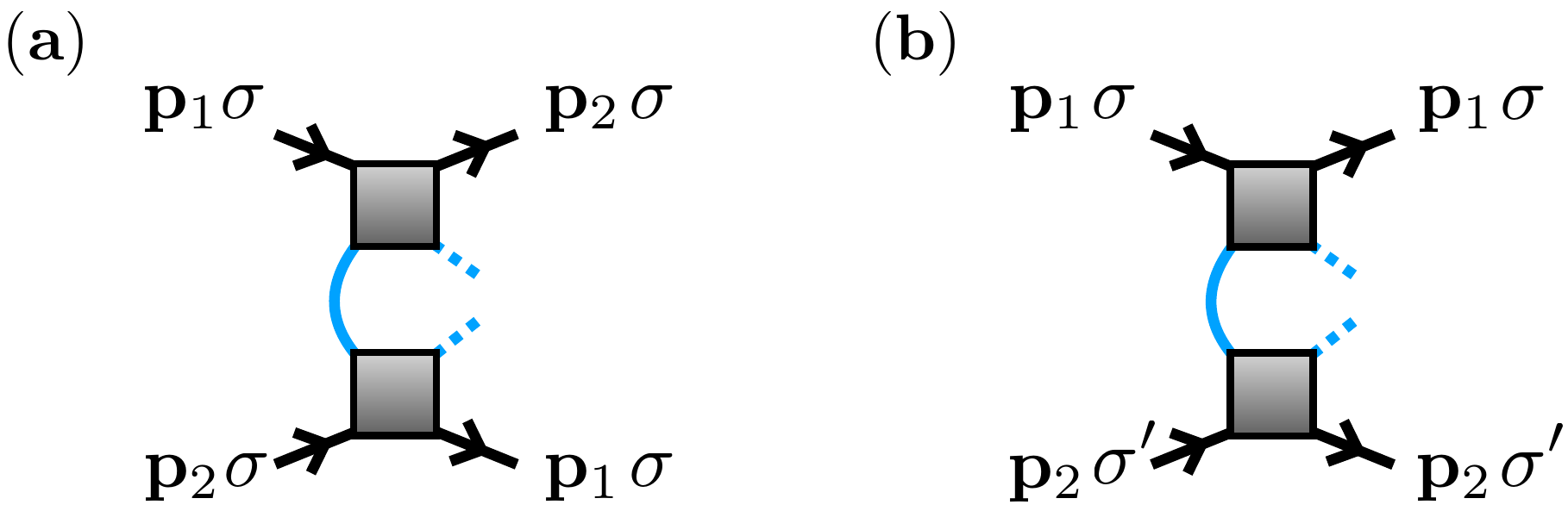}
\caption{(a) Exchange and (b) Hartree diagrams for quasiparticle interactions between Bose polarons at lowest order in perturbation theory. The black lines denote impurity propagators (fermionic, bosonic, or distinguishable), while the blue solid and dotted lines are Bogoliubov excitations and condensate lines of the majority Bose gas, respectively. The squares are the impurity-medium interaction constants, which can in principle be different for distinguishable impurities. The exchange term only exists for fermionic or non-degenerate ($\p_1\neq\p_2$) bosonic impurities, while the Hartree term only contributes to the quasiparticle interactions at fixed medium chemical potential.}
\label{fig:hartreefockBose}
\end{figure}

At this level of perturbation theory, polaron-polaron interactions arise due to the last line of Eq.~\eqref{eq:eomalphasigma}. The origin of these terms is the possibility that one of the two impurities scatters into the momentum occupied by the other impurity via the excitation of a Bogoliubov mode in the medium, with this process being Bose enhanced or Pauli blocked for identical bosonic and fermionic impurities, respectively. Since Bogoliubov modes do not occur at $\k=0$, the process is only present for impurities at different momenta as pointed out in Ref.~\cite{levinsen2025}. This statistical enhancement or suppression is precisely the exchange process discussed in Refs.~\cite{YuPethick2012,Camacho2018}---see Fig.~\ref{fig:hartreefockBose}(a). 

To obtain the polaron interaction strength, we insert Eqs.~\eqref{eq:eomrhosigma} and \eqref{eq:eometasigma} in \eqref{eq:eomalphasigma} to find the total two-polaron energy
\begin{multline}
 E= \epsilon_{\p_1 \sigma}+\epsilon_{\p_2 \sigma'}+  g_{b  \sigma} n+g_{b  \sigma'} n   
   + \frac nV\sum_{\k\neq0} \left(\frac{g_{b  \sigma'}^2 W_{\k}^2}{ E_{\p_1 \sigma;\p_2+\k \sigma';\k}
   } +\frac{g_{b  \sigma}^2W_{\k}^2}{E_{\p_1+\k \sigma;\p_2 \sigma';\k}
   }\right)
  \\ 
      \pm \delta_{ \sigma\sigma'}(1-\delta_{\p_1\p_2}) \frac nV  \left(\frac{  g_{b  \sigma'}^2 W_{\p_1-\p_2}^2}{ E_{\p_1 \sigma;\p_1 \sigma';\p_1-\p_2}
   }
   +\frac{g_{b  \sigma}^2W_{\p_2-\p_1}^2}{E_{\p_2 \sigma;\p_2 \sigma';\p_2-\p_1}
   }\right).
\end{multline}
\new{This equation has the form $E=\Sigma(E)$, where the right-hand side $\Sigma(E)$ is the two-impurity self energy.}
Similarly to the single-polaron problem, \new{within this order of perturbation theory we can} identify $E$ on the right hand side with the leading-order two-polaron energy $E\simeq \epsilon_{\p_1 \sigma}+\epsilon_{\p_2 \sigma'}+g_{b  \sigma} n+g_{b  \sigma'} n$. Rearranging, we then find
\begin{multline}
\label{eq:2BosePolsNoInt}
    E= \epsilon_{\p_1 \sigma} + g_{b  \sigma} n + 
    \frac{n}V\sum_{\k\neq0}
    \frac{g_{b  \sigma}^2W_{\k}^2}{\epsilon_{\p_1 \sigma}-\epsilon_{\p_1+\k \sigma}-E_\k} 
    + \epsilon_{\p_2 \sigma'}
    +g_{b  \sigma'} n +
    \frac{n}V\sum_{\k\neq0}
    \frac{g_{b  \sigma'}^2W_{\k}^2}{\epsilon_{\p_2 \sigma'}-\epsilon_{\p_2+\k \sigma'}-E_\k}\\    
    \mp \delta_{ \sigma\sigma'}(1-\delta_{\p_1\p_2})\frac nV
    \frac{2g_{b \sigma}^2 \epsilon_{\p_1-\p_2 b}}{E_{\p_1-\p_2}^2-(\epsilon_{\p_1 \sigma}-\epsilon_{\p_2 \sigma})^2}\,. 
\end{multline}
The terms in the first line correspond precisely to the (unrenormalized) energies of the individual polarons [see Eq.~\eqref{eq:EpolBose}]. The second line is obtained by using the definition of $W_\k$ and yields the correction due to interactions between the polarons.

By comparing Eq.~\eqref{eq:2BosePolsNoInt}
with the general expression of the two-polaron energy, Eq.~\eqref{eq:Fnp1p2}, we see that the strength of the induced interaction is
\begin{align}\label{eq:inducedinter_gen}
    \new{f_{\p_1\sigma,\p_2\sigma'}}=\mp \delta_{ \sigma\sigma'}(1-\delta_{\p_1\p_2}) \frac{2g_{b \sigma}^2n  \epsilon_{\p_1-\p_2 b}}{E_{\p_1-\p_2}^2-(\epsilon_{\p_1 \sigma}-\epsilon_{\p_2 \sigma})^2}\,.
\end{align}
Thus, we find an induced interaction between indistinguishable impurities $( \sigma=\sigma')$ immersed in a Bose medium which has the opposite sign for bosonic and fermionic impurities in agreement with Refs.~\cite{YuPethick2012,Camacho2018}. By contrast, Eq.~\eqref{eq:inducedinter_gen} shows that, at this level of approximation, there are no medium-induced quasiparticle interactions between distinguishable $(\sigma\neq \sigma')$ impurities or between identical bosonic impurities with equal momenta $\p_1=\p_2$. Finally, we note that the term $(\epsilon_{\p_1 \sigma}-\epsilon_{\p_2 \sigma})$ in the denominator vanishes if we take $|\p_1|=|\p_2|$ or send the impurity mass to infinity. \new{As expected according to our discussion in Section~\ref{sec:QPvsInd}}, in this case $\new{f_{\p_1\sigma,\p_2\sigma'}}$ becomes proportional to the Fourier transform of the Yukawa potential \cite{viverit2000,Naidon2018}. 

Taking the limit of vanishingly small momenta, Eq.~\eqref{eq:inducedinter_gen} gives \new{the quasiparticle interaction}
\begin{align}\label{eq:inducedinter_gen_0k}
F_{n,\sigma\sigma'}
= \mp\delta_{ \sigma\sigma'}(1-\delta_{\p_1\p_2})\frac{g_{ b  \sigma}^2}{g_{bb}}\,.
\end{align}
\new{The presence of a remaining Kronecker delta function highlights the crucial role played by quantum degeneracy for bosonic impurities.}

\new{The quasiparticle interaction in Eq.~\eqref{eq:inducedinter_gen_0k}} diverges in the limit $g_{bb}\to0$, a result of the infinite compressibility of the ideal Bose gas. \new{This scenario, which at the single-impurity level is exactly solvable~\cite{Drescher2021}, gives rise to the bosonic orthogonality catastrophe~\cite{Yoshida2018,Guenther2021,Levinsen2021} and the possibility of Efimov trimers~\cite{Efimov1971,Levinsen2015,Naidon2017}. It is therefore beyond the scope of the perturbative treatment presented in our work.}

\subsubsection{Interacting impurities}
\label{sec:Bose-med_int-imp}
As discussed above, in some cases there is no exchange, and therefore the leading-order medium-induced interaction arises from an altogether different process. Specifically, this is the case for identical, degenerate bosons, or for distinguishable impurities. In both of these cases we find that the leading-order medium-induced \new{quasiparticle} interaction corresponds to a correction to the bare impurity-impurity interaction, e.g., it is what we call a medium-enhanced repulsion in the case of identical bosons~\cite{levinsen2025}. To determine such corrections, we therefore now assume that there is a bare short-range interaction between the impurities. Note that such a short-range interaction does not contribute to the energy in the case of identical fermions with short range interactions due to Fermi statistics. Therefore, the results for fermionic impurities are precisely the same as those obtained above in Eqs.~\eqref{eq:inducedinter_gen} and \eqref{eq:inducedinter_gen_0k}.

To describe the more general case where the bare impurities can interact, we use the following ansatz~\cite{levinsen2025}
\begin{multline}\label{eq:state-ansatzsigmainter} 
    \ket{\Psi}=\Bigg[\alpha_{\p_1\p_2}c^\dag_{\p_1 \sigma}c^\dag_{\p_2 \sigma'}+\sum_{\k\neq0}\rho_{\p_1\p_2\k}c^\dag_{\p_1\sigma}\beta^\dag_{-\k}c^\dag_{\p_2+\k \sigma'}+(1-\delta_{\sigma\sigma'}\delta_{\p_1\p_2})\sum_{\k\neq0}\eta_{\p_1\p_2\k}c^\dag _{\p_1+\k \sigma}\beta^\dag_{-\k}c^\dag_{\p_2\sigma'}
    \\
   +\sum_{\k\neq0}(1-\delta_{\sigma\sigma'}\delta_{\k,\p_2-\p_1})\gamma_{\p_1\p_2\k}c^\dag _{\p_1+\k \sigma}c^\dag_{\p_2-\k \sigma'} \Bigg]\ket{\Phi} \,,
\end{multline}
where the last term allows us to account for the presence of interactions between bare impurities. In the case of indistinguishable impurities $\sigma=\sigma'$, the last sum must exclude $\k=\p_2-\p_1$ (since otherwise the $\gamma$ and $\alpha$ terms are the same), and we furthermore have $\gamma_{\p_1\p_2,\p_2-\p_1-\k}=\gamma_{\p_1\p_2\k}$.

While the presence of bare interactions can render the derivation a bit technical, we can perform a similar calculation as we did in the previous subsection. Doing so, we obtain the following low-momentum \new{$(\p_1,\p_2\rightarrow0)$ quasiparticle interaction constant} for bosonic and distinguishable impurities:
\begin{multline}
\label{eq:inducedinter_gen_0k_wbare}
     F_{n,\sigma\sigma'} = -\delta_{\sigma\sigma'} (1-\delta_{\p_1,\p_2})\frac{g_{ b \sigma}^2}{g_{bb}}    
    + g_{\sigma\sigma'} \left[1+\delta_{\sigma\sigma'} (1-\delta_{\p_1\p_2})\right] \\
   \times \left 
    \{1 +\frac{ 2  g_{b\sigma} g_{b\sigma'}n}{V} \sum_{\k} \left[
 \frac{ W_\k^2}{ \epsilon_{\k \sigma}+\epsilon_{\k \sigma'}}\left(\frac{1}{  E_{\k}+\epsilon_{\k \sigma}}+\frac{1}{  E_{\k}+\epsilon_{\k \sigma'}}\right)
 +  \frac{ W_\k^2}{   \left(  E_{\k}+\epsilon_{\k \sigma'} \right) \left(E_\k+\epsilon_{\k \sigma}\right)} \right]\right\}\, .
\end{multline}
\new{Again, the factor $\delta_{\p_1,\p_2}$ distinguishes degenerate from non-degenerate bosonic impurities.}
By distinguishing the four possible scenarios for impurity statistics and degeneracy, we obtain
\begin{equation}
    F_{n,\sigma\sigma'} 
    = \begin{cases}
    \frac{g_{ b \sigma}^2}{g_{bb}} & \text{fermions} \\
    g_{\sigma\sigma}\left[1+2  g_{b\sigma}^2n\sqrt{\frac{m_b^3}{g_{bb}n}}A(z_\sigma )\right] & \text{bosons }(\p_1=\p_2) \\
    - \frac{g_{ b \sigma}^2}{g_{bb}} +2 g_{\sigma\sigma}\left[1+2  g_{b\sigma}^2n\sqrt{\frac{m_b^3}{g_{bb}n}}A(z_\sigma ) \right] & \text{bosons }(\p_1\neq\p_2) \\
    g_{\up\down} \left[ 1 +2  g_{b\up}g_{b\down}n\sqrt{\frac{m_b^3}{g_{bb}n}}A(z_\up ,z_{\down}) \right] & \text{distinguishable} \,.
    \end{cases}
\end{equation}
Details of this calculation including the generalization to finite impurity momenta are in Appendix~\ref{app:intBose}. The functions $A(z_\sigma )$ and $A(z_\up ,z_{\down})$ with $z_\sigma \equiv m_\sigma /m_b$ take the forms
\begin{subequations}
\begin{align}
\label{eq:A-fun}
    A(z_\sigma ) &=\frac{z_\sigma ^2}{\pi^2(z_\sigma ^2-1)}\left[ 1+(z_\sigma ^2-2) \frac{\arctan(\sqrt{z_\sigma ^2-1})}{\sqrt{z_\sigma ^2-1}}\right]\,, \\
    A(z_\up ,z_{\down}) &=\frac{2z_\up z_\down}{(z_\up^2-z_\down^2)\pi^2}\left[\frac{z_\up^2}{\sqrt{z_\up^2-1}}\arctan(\sqrt{z_\up^2-1})-\frac{z_\down^2}{\sqrt{z_\down^2-1}}\arctan(\sqrt{z_\down^2-1})\right]\,.
\end{align}
\end{subequations}
For distinguishable impurities of the same mass, $m_\up =m_{\down}$, we have $A(z_\up ,z_{\down})=A(z_\sigma )$, and in the particular case where all the masses are equal we have $A(1)=\frac4{3\pi^2}$. The function $A$ is shown in the Appendix~\ref{app:intBose}---see Fig.~\ref{fig:AB_functions}.

\begin{figure}[tbp]    
\centering
\includegraphics[width=.9\linewidth]{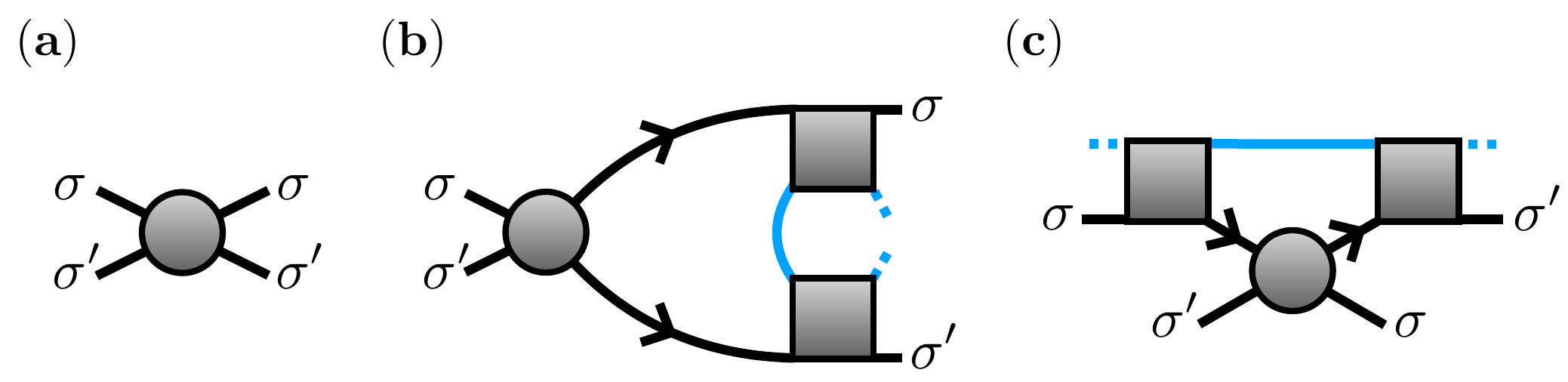}
\caption{(a) Bare impurity interaction (circle) and (b,c) contributions where the medium enhances the bare interactions. Lines and squares as in Fig.~\ref{fig:hartreefockBose}. Note that, in the case of distinguishable impurities ($\sigma \neq \sigma'$), diagram (c) does not include processes involving $g_{\sigma\sigma'}g^2_{b\sigma}$ since these would correspond to a self-energy insertion on the $\sigma$ impurity propagator, and such a self-energy insertion only contributes to the polaron energy at fixed density $n$ and not to the effective interactions.
}
\label{fig:mediumenhancedBose}
\end{figure}

We see from Eq.~\eqref{eq:inducedinter_gen_0k_wbare} that the presence of bare impurity-impurity interactions can play an important role for bosonic and distinguishable impurities. The case of degenerate impurities was recently derived in Ref.~\cite{levinsen2025}. In particular, Ref.~\cite{levinsen2025} demonstrated that the medium-enhancement of the polaron interactions can be related to the Lee-Huang-Yang-type beyond-mean-field energy of the dilute Bose mixture \cite{Larsen63,Petrov2015}, and identified the relevant Feynman diagrams. The diagrams responsible for the polaron-polaron interactions up to second order in the impurity-medium interaction strength are reproduced in Fig.~\ref{fig:mediumenhancedBose}, where the external legs can now be either identical or distinguishable impurities. In the present work, we find that similar medium-induced interactions arise for non-degenerate bosons as well as for distinguishable impurities. In the case of non-degenerate bosons, the terms involving the bare interaction come with a factor of 2 (due to the possibility of exchange) and compete with the term obtained previously, i.e., $-g_{b\sigma}^2/g_{bb}$~\cite{Yu2012,Camacho2018}.

Equation~\eqref{eq:inducedinter_gen_0k_wbare} also clearly shows that the case of distinguishable impurities resembles the case of degenerate bosons, although now the sign of the medium-induced part of the interaction depends on the sign of the product of interaction strengths $g_{\up\down} g_{b\up}g_{b\down}$. In particular, in the case where $m_\up =m_{\down}$ and $g_{b\up}=g_{b\down}$ (such as for $\up$ and $\down$ $^3$He impurities in a $^4$He bath), the two distinguishable impurities can form a fully symmetric wave function. Such a state must have the same ground state energy as that of two indistinguishable bosons, and indeed  by comparing the second and fourth lines of Eq.~\eqref{eq:inducedinter_gen_0k_wbare} we see that this is the case. 

\subsection{Polaron interactions at fixed medium chemical potential}
We now finally wish to determine the induced interaction in the different scenario where the medium chemical potential is fixed rather than the density. At this level of perturbation theory, this is straightforward, since we only need the leading-order number of particles in each polaron dressing cloud. In other words, we can take~\cite{Scazza2022}
\begin{align}
    \Delta N_\sigma =-\pdv{E_{\mathrm{pol},\sigma}}{\mu}\simeq-\frac{g_{b\sigma}}{g_{bb}}\,,
\end{align}
which follows from the mean-field energy $E_{\mathrm{pol},\sigma}=g_{b\sigma}\mu/g_{bb}$. Likewise, for a weakly interacting Bose gas, the density of states at the medium chemical potential is $\mathcal{N}=1/g_{bb}$. Therefore, using Eq.~\eqref{eq:FmuFn2} we find that at second order in perturbation theory
\begin{align}\label{eq:FmuBoseInt}
    F_{\mu,\sigma\sigma'}=F_{n,\sigma\sigma'}-\frac{g_{b\sigma}g_{b\sigma'}}{g_{bb}}\,.
\end{align}
We see that the functional form of the correction at fixed medium chemical potential is the same as the exchange-induced interaction in Eq.~\eqref{eq:inducedinter_gen_0k}. This should not come as a surprise, since this correction corresponds to a Hartree contribution,  Fig.~\ref{fig:hartreefockBose}(b), which in the limit of zero momenta is the same as the exchange contribution,  Fig.~\ref{fig:hartreefockBose}(a). However, note that the Hartree process exists also for distinguishable impurities, as opposed to exchange.

Putting everything together, we find the leading-order low-momentum quasiparticle interactions at fixed chemical potential:
\begin{equation}\label{eq:inducedinter_gen_0k_wbare_mu}
F_{\mu,\sigma\sigma'}
= \begin{cases} 0 & \text{fermions} \\g_{\sigma\sigma}- \frac{g_{ b \sigma}^2}{g_{bb}}
  & \text{bosons }(\p_1=\p_2) 
  \\
        2\left(g_{\sigma\sigma}-\frac{g_{ b \sigma}^2}{g_{bb}}\right)
& \text{bosons }(\p_1\neq\p_2) 
\\ g_{\up\down}- \frac{g_{ b \up}g_{b\down}}{g_{bb}}
 & \text{distinguishable}
    \,.
\end{cases}
\end{equation}
This explicitly illustrates the strongly non-trivial dependence of the polaron interactions on the impurity statistics and degeneracy. Specifically, we find that when the chemical potential of the medium is fixed, identical fermionic impurities do not exhibit induced interactions, while distinguishable or identical bosonic impurities (degenerate or not) do. Furthermore, we again see that, for distinguishable impurities, the sign of the interaction depends on whether the impurity-medium interactions have the same sign for both species. Finally, we see that if the impurities are identical bosons, then there is a statistical enhancement by a factor of 2 for thermal impurities ($\p_1\neq\p_2)$ compared with degenerate impurities ($\p_1=\p_2)$.

\section{Impurities in a Fermi gas}
\label{sec:fermipolaron}
Let us now consider the case of the Fermi polaron where the medium is an ideal Fermi gas. The single-polaron problem has previously been investigated using a host of different techniques. Crucially, it has been shown~\cite{Chevy2006,Combescot2007,Combescot2008} that variational methods based on a single excitation of the medium Fermi sea give excellent agreement with experiment~\cite{Schirotzek2009,Kohstall2012}, even for systems out of equilibrium~\cite{Cetina2016,Parish2016,Liu2019,Adlong2020}. The accuracy of the variational approach has been further supported by diagrammatic Monte Carlo techniques~\cite{Prokofev2008,Houcke2020}. As in the case of the Bose polaron above, the Fermi-polaron problem is very general and occurs also in low-dimensional systems, see, e.g., Refs.~\cite{Zollner2011,Parish2011,Schmidt_PRA2012,Ngampruetikorn2012,Tajima2021}. Beyond ultracold atomic gases, the Fermi-polaron model has also been successfully applied to modeling the optical response of doped atomically thin semiconductors~\cite{Sidler_NatPhys_2017,Efimkin_PRB2017}.

Again we will consider impurities of arbitrary statistics and arbitrary pseudospin $\sigma$. The system is thus described by the general Hamiltonian
\begin{multline}
\label{eq:Ham-Fermi-polaron}
    \hat{H} = \sum_{\k} \epsilon_\k f^\dag_\k f_\k^{} + \sum_{\k,\sigma}\epsilon_{\k \sigma} c_{\k \sigma}^\dag c_{\k \sigma}^{} + \sum_\sigma  \frac{g_{f\sigma}}{V} \sum_{\k\k'\q} c^\dag_{\k \sigma} f^\dag_{\k'} f_{\k'+\q}^{} c_{\k-\q \sigma}^{} \\ 
    + \sum_{\sigma\sigma'}\frac{g_{\sigma\sigma'}}{2V} \sum_{\k\k'\q} c^\dag_{\k \sigma} c^\dag_{\k'\sigma'} c_{\k'+\q \sigma'} c_{\k-\q \sigma} \, ,
\end{multline}
where $f_\k^{\dag}$ ($f_\k^{}$) are Fermi creation (annihilation) operators of the medium with momentum $\k$ and energy $\epsilon_{\k} = k^2/2m_f$. The impurities are described exactly as in Section~\ref{sec:bosepolaron}, while $g_{f\sigma}$ is the fermion-impurity interaction, which is related to the scattering length $a_{f\sigma}$ by $g_{f\sigma} = 2\pi a_{f\sigma}/m_{f\sigma}$, where  $m_{f\sigma}=m_fm_\sigma /(m_f+m_\sigma )$ is the impurity-fermion reduced mass. 

\subsection{Single polaron problem}
\label{sec:single-F-polaron}
The Fermi polaron at momentum $\p$ is well described via Chevy's ansatz~\cite{Chevy2006}
\begin{equation}
    \ket{\Psi} = \left(\alpha_\p c_{\p \sigma}^\dag + \sum_{\k,\q} \alpha_{\p \k \q} c_{\p+\q-\k \sigma}^\dag f_\k^\dag f_\q^{}\right) \fs \, ,
\end{equation}
which assumes that the impurity is dressed by at most a single particle-hole excitation of the Fermi sea. Correspondingly, we have $k>k_F \geq q$ here and everywhere in the following, where $k_F$ is the Fermi momentum. The Fermi sea $\fs  = \prod_{\q} f_\q^\dag \ket{0}$ is further characterized by the Fermi energy $E_F=\frac{k_F^2}{2m_f}$ and density $n = \frac{k_F^3}{6\pi^2}$.

We evaluate the equations of motion like in the case of a bosonic medium in Section~\ref{sec:bosepolaron}, which gives
\begin{subequations}
\label{eq:1F-polaron}
\begin{align}
\label{eq:1F-polaron1}
    \left(E- \epsilon_{\p \sigma} - g_{f\sigma}n\right)\alpha_\p &=  \frac{g_{f\sigma}}{V} \sum_{\k\q} \alpha_{\p\k\q} \\
    \left(E - E_{\p\k\q \sigma} - g_{f\sigma} n\right) \alpha_{\p\k\q} &= \frac{g_{f\sigma}}{V} \alpha_\p +  \frac{g_{f\sigma}}{V} \sum_{\k'}\alpha_{\p\k'\q}
+ \frac{g_{f\sigma}}{V} \sum_{\q'} \alpha_{\p\k\q'}\; ,
\label{eq:1F-polaron2}
\end{align}
\end{subequations}
where $E_{\p\k\q \sigma} = \epsilon_{\p+\q-\k \sigma} + \epsilon_\k - \epsilon_\q$ and where we have measured the energy from that of the single-component Fermi gas,
$E_\mathrm{FS} = \sum_{\q} \epsilon_\q = \frac{3}{5} E_F V n$.

The polaron energy can be obtained perturbatively up to second order in  $g_{f\sigma}$ by inserting~\eqref{eq:1F-polaron2} into~\eqref{eq:1F-polaron1}, giving:
\begin{equation}
    E \simeq \epsilon_{\p \sigma} + g_{f\sigma}n +  \frac{g_{f\sigma}^2}{V^2} \sum_{\k \q}\frac{1}{E - E_{\p\k\q \sigma} - g_{f\sigma} n}\; .
\end{equation}
One then substitutes $E \simeq \epsilon_{\p \sigma}+g_{f\sigma}n$ on the right-hand side to find
\begin{equation}
\label{eq:un-ren-Fpol}
    E \simeq \epsilon_{\p \sigma} + g_{f\sigma}n +  \frac{g_{f\sigma}^2}{V^2} \sum_{\k \q}\frac{1}{\epsilon_{\p \sigma} - E_{\p\k\q \sigma}}\; .
\end{equation}
Here, the divergence of the sum on $\k$ can be regularized by the replacement $g_{f\sigma} \mapsto g_{f\sigma} + (g_{f\sigma}^2/V) (\sum_\q 1/\bar{\epsilon}_{\q\sigma} + \sum_\k 1/\bar{\epsilon}_{\k\sigma})$, 
with $\bar{\epsilon}_{\k\sigma} = \epsilon_{\k \sigma} + \epsilon_\k$. The final expression of the Fermi-polaron energy at momentum $\p$ is:
\begin{equation}
\label{eq:1-Fermi-polaron-pert}
    E_{\mathrm{pol},\sigma} (\p) \simeq \epsilon_{\p \sigma} + g_{f\sigma} n \left(1 + \frac{g_{f\sigma}}{V} \sum_\q \frac{1}{\bar{\epsilon}_{\q\sigma}} \right)
    +\frac{g_{f\sigma}^2}{V^2} \sum_{\k \q} \left(\frac{1}{\epsilon_{\p \sigma} - E_{\p\k\q \sigma}} + \frac{1}{\bar{\epsilon}_{\k\sigma}}\right)\; .
\end{equation}
The polaron energy at weak interactions was first obtained in Ref.~\cite{Bishop1973}. In particular, when $m_\sigma=m_f$ we find $E_{\mathrm{pol},\sigma}(0)=g_{f\sigma}n(1+\frac3{2\pi}k_Fa_{f\sigma})$.

\subsection{Polaron interactions at fixed medium density}

\subsubsection{Non-interacting impurities}
As in the case of the Bose polaron in Section~\ref{sec:bosepolaron}, we begin by investigating the polaron interactions in the absence of any bare interaction between impurities. To this end, we introduce a two-impurity variational ansatz featuring a single excitation of the Fermi sea:
\begin{multline}
\label{eq:2-pol-F-med}
    \ket{\Psi} = \left( \alpha_{\p_1\p_2} c^{\dag}_{\p_1 \sigma} c^{\dag}_{\p_2 \sigma'} +
  \sum_{\k\q} \rho_{\p_1\p_2\k\q}
  c^{\dag}_{\p_1\sigma} f^{\dag}_{\k} f_{\q}^{} c^{\dag}_{\p_2 -\k+\q \sigma'} \right.\\ \left.
   + (1-\delta_{\p_1\p_2} \delta_{\sigma\sigma'}) \sum_{\k\q} \eta_{\p_1\p_2\k\q}
  c^{\dag}_{\p_1 -\k+\q \sigma} f^{\dag}_{\k} f_{\q}^{} c^{\dag}_{\p_2 \sigma'} \right) \fs  \; .
\end{multline}
Again, the impurities of species $\sigma$ and $\sigma'$ are at momenta $\p_1$ and $\p_2$, respectively, before scattering with the medium. We assume that $\p_1\ne \p_2$ for identical ($\sigma=\sigma'$) fermionic impurities, while for identical bosonic impurities with equal momenta the $\eta$-term is absent. The normalization of $\ket{\Psi}$ requires
\begin{multline}
\label{eq:2-pol-F-med-nor}
    1 = \bra{\Psi} \ket{\Psi} = (1+\delta_{\p_1\p_2} \delta_{\sigma\sigma'}) |\alpha_{\p_1\p_2}|^2 + \sum_{\k\q} |\rho_{\p_1\p_2\k\q}|^2 + (1-\delta_{\p_1\p_2} \delta_{\sigma\sigma'}) \sum_{\k\q} |\eta_{\p_1\p_2\k\q}|^2 \\
    \pm \delta_{\sigma\sigma'} (1-\delta_{\p_1\p_2}) \sum_{\k\q} \left(\delta_{\q-\k , \p_1-\p_2}|\rho_{\p_1\p_2\k\q}|^2 + \delta_{\q-\k , \p_2-\p_1}|\eta_{\p_1\p_2\k\q}|^2\right)\; ,
\end{multline}
where $\pm$ refers to bosonic and fermionic identical impurities, respectively. Note that the scenario where one impurity is scattered into the same momentum as the other impurity via a particle-hole excitation of the Fermi gas is absent if $\p_1=\p_2$ since then $\q-\k = \pm (\p_1 - \p_2)=0$ can never be satisfied.

The equations of motion are obtained as before, by evaluating $\partial_{\lambda*}\bra{\Psi} (E-\hat{H}-E_\mathrm{FS}) \ket{\Psi}=0$ with $\lambda\in\{ \alpha_{\p_1\p_2},\rho_{\p_1\p_2\k \q},\eta_{\p_1\p_2\k\q }\}$:
\begin{subequations}
\label{eq:Fermi-med-eq}
\begin{align}
    E \alpha_{\p_1\p_2} =& \left(\epsilon_{\p_1 \sigma} + \epsilon_{\p_2 \sigma'} + g_{f\sigma} n + g_{f\sigma'} n\right)\alpha_{\p_1\p_2} + \frac{g_{f\sigma'}}{V} \sum_{\k\q} \rho_{\p_1\p_2 \k\q} \nn \\ 
    &+ (1-\delta_{\p_1\p_2} \delta_{\sigma\sigma'}) \frac{g_{f\sigma}}{V} \sum_{\k\q} \eta_{\p_1\p_2 \k\q}  \nonumber\\
    &\pm \delta_{ \sigma\sigma'}(1-\delta_{\p_1\p_2}) \frac{g_{f\sigma}}{V} \sum_{\k\q} \left(\delta_{\q-\k , \p_1 - \p_2} \rho_{\p_1\p_2 \k\q} 
      + \delta_{\q-\k , \p_2 - \p_1} \eta_{\p_1\p_2 \k\q}\right) \label{eq:Fermi-med-eq-a}\\
    E \rho_{\p_1\p_2 \k\q} =& \left(\epsilon_{\p_1 \sigma} + E_{\p_2\k\q \sigma'} + g_{f\sigma} n +  g_{f\sigma'} n\right)\rho_{\p_1\p_2 \k\q}  + (1+\delta_{\p_1\p_2} \delta_{\sigma\sigma'}) \frac{g_{f\sigma'}}{V} \alpha_{\p_1\p_2}\label{eq:Fermi-med-eq-b}\\
    E \eta_{\p_1\p_2 \k\q} =& \left(\epsilon_{\p_2 \sigma'} + E_{\p_1\k\q \sigma} + g_{f\sigma} n + g_{f\sigma'} n \right)\eta_{\p_1\p_2 \k\q}   + \frac{g_{f\sigma}}{V} \alpha_{\p_1\p_2}\label{eq:Fermi-med-eq-c}\; ,
\end{align}
\end{subequations}
where, evidently, Eq.~\eqref{eq:Fermi-med-eq-c} is absent for identical bosons with $\p_1 = \p_2$.\footnote{In Eq.~\eqref{eq:Fermi-med-eq}, we are neglecting terms that are $O(V^{-2})$. We also neglect the following terms featuring repeated scattering with particle-hole excitations because these are small in the perturbative limit: $- \frac{g_{f\sigma'}}{V} \sum_{\q'} \rho_{\p_1\p_2 \k\q'}$ and $+ \frac{g_{f\sigma'}}{V} \sum_{\k'} \rho_{\p_1\p_2 \k'\q}$ in Eq.~\eqref{eq:Fermi-med-eq-b} and $- \frac{g_{f\sigma}}{V} \sum_{\q'} \eta_{\p_1\p_2 \k\q'}$ and $+ \frac{g_{f\sigma}}{V} \sum_{\k'} \eta_{\p_1\p_2 \k'\q}$ in Eq.~\eqref{eq:Fermi-med-eq-c}.} At this level of perturbation theory, the effective polaron-polaron interactions originate from the last line in Eq.~\eqref{eq:Fermi-med-eq-a} and are due to statistical effects where the scattering of two identical impurities into the same momentum is enhanced or suppressed for bosonic and fermionic impurities, respectively. 

As for the Bose medium, by substituting Eqs.~\eqref{eq:Fermi-med-eq-b} and~\eqref{eq:Fermi-med-eq-c} into~\eqref{eq:Fermi-med-eq-a}, we obtain a closed equation for the total two-polaron energy:
\begin{multline}
    E = \epsilon_{\p_1 \sigma} + \epsilon_{\p_2 \sigma'} + g_{f\sigma} n + g_{f\sigma'} n \\
    + \frac{1}{V^2} \sum_{\k\q} \left[\frac{g_{f\sigma'}^2}{E - (\epsilon_{\p_1 \sigma} + E_{\p_2\k\q \sigma'} + g_{f\sigma} n + g_{f\sigma'} n)} 
    + \frac{g_{f\sigma}^2}{E - (\epsilon_{\p_2 \sigma'} + E_{\p_1\k\q \sigma} + g_{f\sigma} n + g_{f\sigma'} n)} \right]   
    \\
    \pm \delta_{ \sigma\sigma'}(1-\delta_{\p_1\p_2}) \frac{g_{f\sigma}^2}{V^2} \sum_{\k} \left[\frac{1}{E - (2 \epsilon_{\p_1 \sigma} + \epsilon_{\k} - \epsilon_{\k + \p_1 - \p_2} + 2g_{f\sigma} n)}\right.\\
    \left.+ \frac{1}{E - (2 \epsilon_{\p_2 \sigma} + \epsilon_{\k} - \epsilon_{\k + \p_2 - \p_1} + 2g_{f\sigma} n)}\right]
    \; .
\end{multline}
This equation can be solved perturbatively by substituting the mean-field energy of both polarons, $E \simeq \epsilon_{\p_1 \sigma} + \epsilon_{\p_2 \sigma'} + g_{f\sigma} n + g_{f\sigma'} n$, on the r.h.s, giving:
\begin{multline}
    E = \left(\epsilon_{\p_1 \sigma} + g_{f\sigma}n +  \frac{g_{f\sigma}^2}{V^2} \sum_{\k \q}\frac{1}{\epsilon_{\p_1 \sigma}- E_{\p_1\k\q \sigma}}\right) + \left(\epsilon_{\p_2 \sigma'} + g_{f\sigma'}n +  \frac{g_{f\sigma'}^2}{V^2} \sum_{\k \q}\frac{1}{\epsilon_{\p_2 \sigma'}- E_{\p_2\k\q \sigma'}}\right)\\
    \pm \delta_{ \sigma\sigma'}(1-\delta_{\p_1\p_2}) \frac{g_{f\sigma}^2}{V^2} \sum_{\k} \left[\frac{1}{\epsilon_{\p_2 \sigma} - \epsilon_{\p_1 \sigma} + \epsilon_{\k + \p_1 - \p_2}
    - \epsilon_{\k}} + \frac{1}{\epsilon_{\p_1 \sigma} - \epsilon_{\p_2 \sigma} + \epsilon_{\k + \p_2 - \p_1}
    - \epsilon_{\k}}\right]
    \; .
\end{multline}
%

The first two terms in brackets in the first line correspond to the (unrenormalized) energies of the individual polarons given in Eq.~\eqref{eq:un-ren-Fpol}. Thus, the leading-order correction to the two-polaron energy---and, therefore, to the mediated quasiparticle interaction constant~\eqref{eq:Fnp1p2}---can be readily obtained
\begin{align}
\label{eq:induced-inter_Fermi-med}
    \new{f_{\p_1\sigma,\p_2\sigma'}} &= \mp\delta_{ \sigma\sigma'}(1-\delta_{\p_1\p_2}) g_{f\sigma}^2 L(\p_1-\p_2, \epsilon_{\p_2 \sigma} - \epsilon_{\p_1 \sigma})\; ,
\end{align}
in terms of the zero-temperature dynamic Lindhard screening function 
\begin{align}
    L(\q, \omega) &= \frac{1}{V} \sum_\k \frac{n_\k - n_{\k+\q}}{\omega + \epsilon_{\k + \q} - \epsilon_{\k}}\;,
\end{align}
which appears in the theory of electron gases and can be evaluated analytically at zero temperature~\cite{Giuliani-Vignale-book2008}.\footnote{We use the sign convention of Ref.~\cite{PethickSmithBook} for defining the Lindhard screening function, which has a minus sign compared to Ref.~\cite{Giuliani-Vignale-book2008}.} Similarly to the Bose polaron case, the term $(\epsilon_{\p_1 \sigma}-\epsilon_{\p_2 \sigma})$ in Eq.~\eqref{eq:induced-inter_Fermi-med} vanishes if we take $|\p_1|=|\p_2|$ or send the impurity mass to infinity. In this case, the zero-temperature Lindhard function, and thus the Fermi polaron interaction \new{$f_{\p_1\sigma,\p_2\sigma'}$,} 
becomes proportional to the Fourier transform of the \new{RKKY} interaction potential~\cite{Nishida2009,Giuliani-Vignale-book2008}.

\begin{figure}[tbp]    
\centering
\includegraphics[width=.65\linewidth]{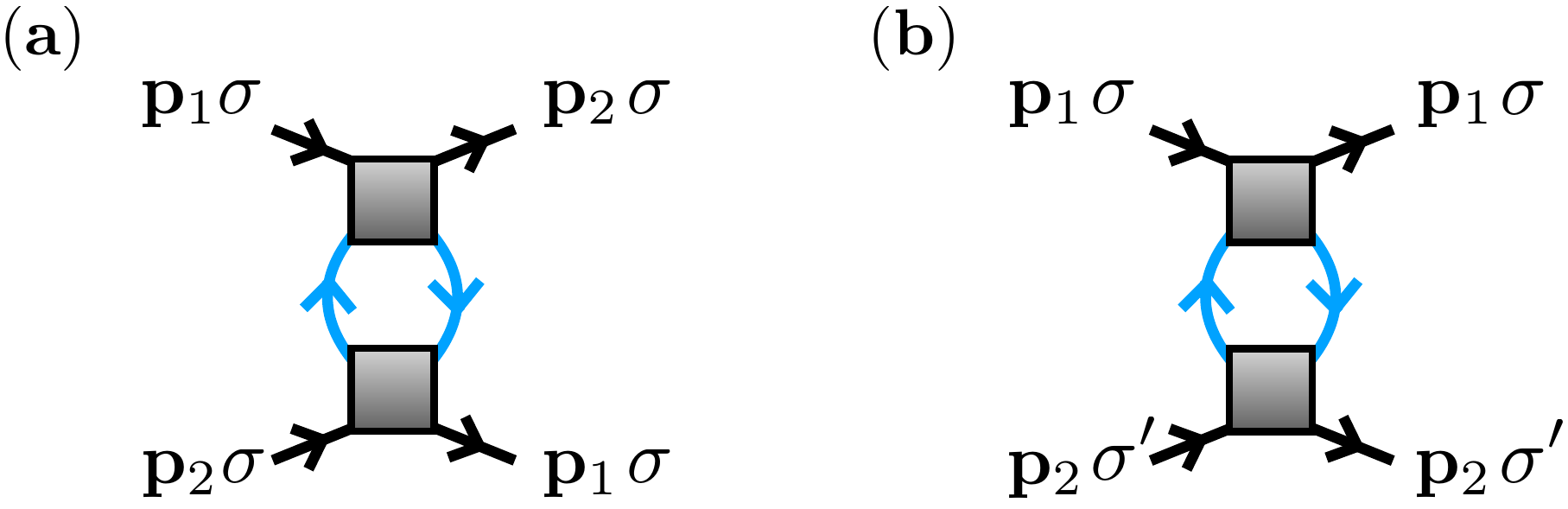}
\caption{(a) Exchange and (b) Hartree diagrams for quasiparticle interactions between Fermi polarons at lowest order in perturbation theory. The black lines denote fermionic, bosonic, or distinguishable impurities, while the blue lines denote majority particle propagators. The squares are the impurity-medium interaction constants, and can in principle be different for distinguishable impurities. As in Fig.~\ref{fig:hartreefockBose}, the exchange term only exists for fermionic or non-degenerate bosonic impurities. The Hartree term only contributes to the quasiparticle interactions at fixed medium chemical potential, not at fixed medium density.}
\label{fig:hartreefockFermi}
\end{figure}

\new{Taking} the static $\omega \to 0$ and \new{then the} long wavelength $\q \to \0$ limits, the Lindhard function recovers the density of states at the Fermi surface $\mathcal{N} (E_F)= \frac{3 n}{2 E_F}$.
Thus, in the limit of vanishingly small momenta, \new{$\p_1,\p_2\to0$},
\begin{equation}
\label{eq:induced-inter_Fermi-med2}
    F_{n,\sigma\sigma'}
    = \mp\delta_{ \sigma\sigma'}(1-\delta_{\p_1\p_2}) \frac{3 n g_{f\sigma}^2}{2 E_F} \; .
\end{equation}
The results for fermionic and (non-degenerate) bosonic impurities in a Fermi sea coincide with those derived in Refs.~\cite{YuPethick2012} and~\cite{Scazza2022}, respectively, by considering the process where two impurities are exchanged---see Fig.~\ref{fig:hartreefockFermi}(a). Equation \eqref{eq:induced-inter_Fermi-med2} explicitly demonstrates that, at this level of perturbation theory, there are no interactions between identical bosonic impurities at the same momentum, or between distinguishable particles. We note that we expect there to be contributions to the quasiparticle interactions at higher order in both of these scenarios, even in the absence of bare impurity-impurity interactions, similarly to what was found for impurities in a Bose medium~\cite{levinsen2025}.

\subsubsection{Interacting impurities}
As for the case of a Bose medium (Section~\ref{sec:Bose-med_int-imp}), we now summarize the results for the corrections to the bare impurity-impurity interaction induced by the medium. We thus allow the impurities to interact via a bare contact term $g_{\sigma\sigma'}$, as in~\eqref{eq:Ham-Fermi-polaron}, and consider the following variational ansatz, where, in addition to the terms in Eq.~\eqref{eq:2-pol-F-med}, we include the possibility for impurities to scatter into different momentum states:
\begin{multline}
\label{eq:2-pol-F-med-int}
  \ket{\Psi} = \left[ \alpha_{\p_1\p_2} c^{\dag}_{\p_1 \sigma} c^{\dag}_{\p_2 \sigma'} +
  \sum_{\k\q} \rho_{\p_1\p_2\k\q}
  c^{\dag}_{\p_1\sigma} f^{\dag}_{\k} f_{\q}^{} c^{\dag}_{\p_2 -\k+\q \sigma'}\right. \\
  \left. + (1-\delta_{\p_1\p_2} \delta_{\sigma\sigma'}) \sum_{\k\q} \eta_{\p_1\p_2\k\q}
  c^{\dag}_{\p_1 -\k+\q \sigma} f^{\dag}_{\k} f_{\q}^{} c^{\dag}_{\p_2 \sigma'} \right.\\
    \left. + \sum_{\K 
    \neq0}(1-\delta_{\sigma\sigma'}\delta_{\K,\p_2-\p_1})\gamma_{\p_1\p_2\K}c^\dag _{\p_1+\K \sigma}c^\dag_{\p_2-\K \sigma'}  \right]\fs  \; .
\end{multline}
As in the case of non-interacting impurities, we require $\p_1 \ne \p_2$ for identical ($\sigma = \sigma'$) fermionic impurities, and we take $k > k_F \geq q$, while $\K$ is unconstrained by the Fermi sea. For indistinguishable bosonic impurities, the last sum in Eq.~\eqref{eq:2-pol-F-med-int} excludes the contribution from $\K=\p_2 - \p_1$ in order to avoid redundancy with the first term. Also, in this case, the variational function $\gamma_{\p_1\p_2\K}$ must satisfy the symmetry condition $\gamma_{\p_1\p_2, \p_2 - \p_1 -\K} = \gamma_{\p_1\p_2, \K}$. 

The calculation proceeds along the same lines as in the previous cases, and the details are provided in Appendix~\ref{app:intFermi}. As discussed earlier, Fermi statistics forbid identical fermions from interacting through a short-range potential, so that for identical fermionic impurities one simply has that $\lim_{\p_1,\p_2 \to \0} \new{f_{\p_1\sigma,\p_2\sigma}} = 3 n g_{f\sigma}^2/2 E_F$ as in Eq.~\eqref{eq:induced-inter_Fermi-med2}. For the other cases, we obtain the following perturbative results for the low-momentum \new{($\p_1,\p_2 \to \0$) quasiparticle} interactions:
\begin{multline}
F_{n,\sigma\sigma'}
= -\delta_{ \sigma\sigma'}(1-\delta_{\p_1\p_2}) \frac{3 n g_{f\sigma}^2}{2 E_F}\\
    + g_{\sigma\sigma'} \left[1+\delta_{\sigma\sigma'}(1-\delta_{\p_1,\p_2})\right]
    \left\{1 + 2 \frac{g_{f\sigma} g_{f\sigma'}}{V^2} \sum_{\k\q}\left[\frac{1}{\epsilon_{\q-\k \sigma} +\epsilon_{\q-\k \sigma'}} \left(\frac{1}{\epsilon_{\q-\k \sigma} +\epsilon_\k -\epsilon_\q}\right.\right.\right.\\
    \left.\left.\left. + \frac{1}{\epsilon_{\q-\k \sigma'} +\epsilon_\k -\epsilon_\q}\right) + \frac{1}{(\epsilon_{\q-\k \sigma} +\epsilon_\k -\epsilon_\q)(\epsilon_{\q-\k \sigma'} +\epsilon_\k -\epsilon_\q)}\right]\right\}\; .
\end{multline}
Thus, considering the four possible impurity scenarios, we find that:
\begin{equation}
\label{eq:inducedinterfermioninterimp}
F_{n,\sigma\sigma'}
= \begin{cases}
        \frac{3 n}{2 E_F}g_{f\sigma}^2 &\text{fermions}\\
    g_{\sigma\sigma}\left[1 + 9 \left(\frac{g_{f\sigma} n}{E_F}\right)^2B(z_\sigma )\right] &\text{bosons}\ (\p_1 = \p_2)\\
    -\frac{3n}{2E_F} g_{f\sigma}^2 + 2 g_{\sigma\sigma}\left[1 + 9 \left(\frac{g_{f\sigma} n}{E_F}\right)^2B(z_\sigma )\right] &\text{bosons}\ (\p_1 \ne \p_2)\\
    g_{\up\down}\left[1 + 9 g_{f\up} g_{f\down} \left(\frac{n}{E_F}\right)^2 B(z_{\up},z_{\down})\right] &\text{distinguishable}
    \end{cases}\; ,
\end{equation}
where the functions $B(z_\up,z_{\down})$  and $B(z_\sigma)$ can be found in Appendix~\ref{app:intFermi} and in Fig.~\ref{fig:AB_functions}. For impurities with equal masses $B(z_\uparrow,z_\downarrow) = B(z_\sigma)$, while for $m_f=m_\uparrow=m_\downarrow$, we have $B(1) = B(1,1)=1/4 + \pi^2/16$. 

\begin{figure}[tbp]    
\centering
\includegraphics[width=.9\linewidth]{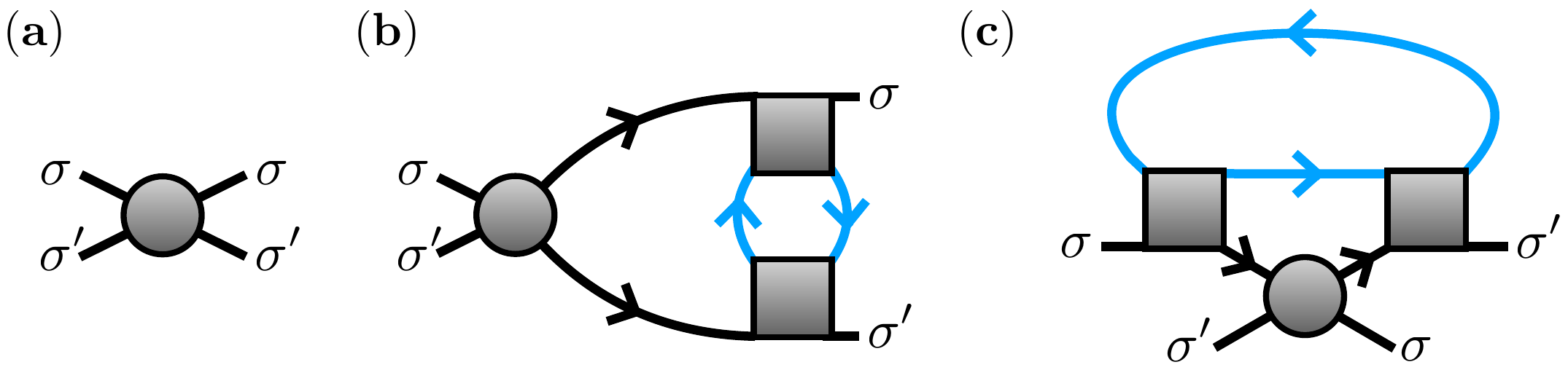}
\caption{(a) Bare impurity interaction (circle) and (b,c) medium-enhanced polaron interactions. Lines and squares as in Fig.~\ref{fig:hartreefockFermi}. Note that, in the case of distinguishable impurities ($\sigma \neq \sigma'$), diagram (c) does not include processes involving $g_{\sigma\sigma'}g^2_{f\sigma}$ since these would correspond to self-energy insertions on the $\sigma$ impurity propagator, which do not contribute to effective polaron-polaron interactions at fixed density $n$.}
\label{fig:mediumenhancedFermi}
\end{figure}

We find that the presence of bare interactions can play an important role in the quasiparticle interactions, similarly to the Bose-polaron case. Specifically, for bosonic impurities the bare impurity-impurity interactions are effectively enhanced by the presence of the medium, while the sign of the medium-induced part of the interaction for distinguishable impurities depends on the sign of $g_{\up\down} g_{f\up}g_{f\down}$. The modifications of the bare interactions arise from quantum fluctuations of the Fermi gas, with the associated diagrams shown in Fig.~\ref{fig:mediumenhancedFermi}.

\subsection{Polaron interactions at fixed medium chemical potential}

To instead obtain the polaron interactions at fixed $\mu$, we need the number of particles in the dressing cloud. At leading order in perturbation theory, this is
\begin{equation}
    \Delta N_\sigma=-\pdv{E_{\mathrm{pol},\sigma}}{\mu}=-\frac32\frac{g_{f\sigma}n}{E_F}\;,
\end{equation}
where both $E_F=\mu$ and $n=(2mE_F)^{3/2}/6\pi^2$ are functions of the medium chemical potential. Similarly, the density of states defined in Eq.~\eqref{eq:N} is $\mathcal{N}=\frac32n/E_F$. 

Thus, we find the leading-order result for the low-momentum quasiparticle interactions
\begin{equation} \label{eq:inducedFermipolmu}
  F_{\mu,\sigma\sigma'}
  = 
  \begin{cases} 0 & \text{fermions} \\ g_{\sigma\sigma}-\frac{3 n}{2 E_F}g_{f\sigma}^2
  & \text{bosons }(\p_1=\p_2) 
  \\
        2\left(g_{\sigma\sigma}-\frac{3 n}{2E_F}g_{f\sigma}^2\right)
& \text{bosons }(\p_1\neq\p_2) 
\\ g_{\up\down}- \frac{3 n}{2E_F}g_{f\up}g_{f\down}
 & \text{distinguishable}
     \,.
\end{cases}
\end{equation}
By comparing with Eq.~\eqref{eq:inducedinter_gen_0k_wbare_mu}, we see that Eq.~\eqref{eq:inducedFermipolmu} bears a lot of similarity with the corresponding result for Bose polarons.

The factor 2 enhancement of the interaction energy for non-degenerate bosons compared with degenerate bosons has not yet been observed. However, we note that the experiment in Ref.~\cite{DeSalvo2019} observed fermion-mediated interactions between bosonic atoms, with the interaction extracted from the change in the profile of the trapped bosons due to the presence of the medium. In other words, the experiment should be well described via the local density approximation such that it was, effectively, conducted at fixed medium chemical potential. The authors found that the magnitude of the interaction was enhanced by a factor 1.7 compared with what would be observed if the boson impurities were all in a BEC. We can therefore speculate that this enhancement was, at least in part, due to thermal depletion of the impurity condensate.

\section{Conclusion and outlook}
\label{sec:conc}
To conclude, we have demonstrated that the medium-induced interactions between polarons depend sensitively on the nature of the impurities and on the constraints on how the medium can respond to a perturbation. As we have discussed, these results have important implications for the interpretation of current and future experiments on induced interactions, both in cold atomic gases and in atomically thin semiconductors. The observation of some of the effects predicted here---such as the possibility of a medium-enhanced repulsion between impurities, or the statistical enhancement of interactions between thermal rather than degenerate impurities---appears well within experimental reach. In particular, a recent experiment~\cite{DeSalvo2019} observed fermion-mediated interactions between bosonic impurities which were larger than those expected for an impurity BEC by a factor 1.7. Our results imply that this could, at least in part, be explained by thermal depletion of the impurity condensate.

To provide specific predictions for Bose and Fermi polarons of arbitrary statistics, we formulated a wave function approach that directly allowed us to extract the polaron interactions in the perturbative weak-coupling regime. While we focused on 3D gases, our approach is very general and it can therefore be applied to a host of other scenarios, such as low-dimensional systems, exciton impurities in semiconductors, protons in neutron stars, impurities in neutron matter, and polarons living on a lattice. In the future, it would be interesting to push our approach beyond the perturbative regime where it is likely to form the basis of theories that can incorporate multiple constraints on the medium while simultaneously capturing the whole range of predicted contributions to polaron interactions, namely exchange effects, medium-enhanced interactions, and phase-space filling. In particular, there is the possibility of medium-induced pairing and bipolaron bound states, which are fundamentally different from the unbound scattering polarons considered in this work. 

\section*{Acknowledgements}
We gratefully acknowledge fruitful discussions with Cosetta Baroni, Georg Bruun, Fr{\'e}d{\'e}ric Chevy, Tilman Enss, Victor Gurarie, Pietro Massignan, Chris Pethick, Matteo Zaccanti, and Martin Zwierlein.
JL, OB, and MMP acknowledge support from the Australian Research Council ARC Centre of Excellence in Future Low-Energy Electronics Technologies (CE170100039). 
JL and MMP are supported through Australian Research Council Discovery Project DP240100569 and Future Fellowship FT200100619, respectively. OB also acknowledges support from the Deutsche Forschungsgemeinschaft (DFG) via the Collaborative Research Centre SFB 1225 ISOQUANT (Project-ID No. 273811115). FMM acknowledges financial support from the Spanish Ministry of Science, Innovation and Universities through the ``Maria de Maetzu'' Programme for Units of Excellence in R\&D (CEX2023-001316-M), and from the Ministry of Science, Innovation and Universities MCIN/AEI/10.13039/501100011033, FEDER UE,  project No.~PID2023-150420NB-C31 (Q).

\appendix
\renewcommand{\theequation}{\Alph{section}\arabic{equation}}
\setcounter{equation}{0}

\section{Interacting Bose polarons}
\label{app:intBose}
In this appendix, we provide the detailed derivation of the Bose polaron interactions in the presence of interactions between bare impurities given in Eq.~\eqref{eq:inducedinter_gen_0k_wbare} of the main text. For the general ansatz given in Eq.~\eqref{eq:state-ansatzsigmainter}, the norm is
\begin{multline}
\label{eq:norm2impint}
1=\braket{\Psi}{\Psi}=(1+\delta_{\sigma\sigma'}\delta_{\p_1\p_2})|\alpha_{\p_1\p_2}|^2 +\sum_{\k\neq0}|\rho_{\p_1\p_2\k}|^2 + (1-\delta_{\sigma\sigma'}\delta_{\p_1\p_2})\sum_{\k\neq0}|\eta_{\p_1\p_2\k}|^2 \\  + \sum_{\k\neq0}|\gamma_{\p_1\p_2\k}|^2\left(1\pm \delta_{\sigma\sigma'}\right) \left(1- \delta_{\sigma\sigma'}\delta_{\k,\p_2-\p_1}\right)
\\
\pm \delta_{\sigma\sigma'} (1-\delta_{\p_1\p_2}) \left(|\rho_{\p_1\p_2,\p_1-\p_2}|^2 + |\eta_{\p_1\p_2,\p_2-\p_1}|^2\right)\, .
\end{multline}
In order to keep the derivation of the polaron interactions clear, in the following we consider three different cases separately.

\subsection{Indistinguishable bosonic impurities with equal momenta}
For indistinguishable bosonic impurities at equal momenta $\p_1=\p_2=\p$, the ansatz simplifies to
\begin{align}
    \ket{\Psi}=&\left[\alpha_{\p}c^\dag_{\p \sigma}c^\dag_{\p \sigma}+\sum_{\k\neq0}\rho_{\p \k}c^\dag_{\p \sigma}\beta^\dag_{-\k}c^\dag_{\p+\k \sigma}+\sum_{\k\neq0} \gamma_{\p \k}c^\dag _{\p+\k \sigma}c^\dag_{\p-\k \sigma} \right]\ket{\Phi} \label{eq:ansatzinter_ind_equalp}
\end{align}
Using $W_{\k}=W_{-\k}$ , $\gamma_{\p, \k}=\gamma_{\p, -\k}$, we obtain the equations of motion
\begin{subequations} 
\begin{align} \label{eq:equala}
   \left(E_{\p\sigma;\p\sigma}-\frac{g_{\sigma\sigma}}{V}\right)\alpha_{\p} &=\frac{g_{b \sigma}\sqrt{N}}{V}\sum_{\k}W_{\k}\rho_{\p \k}+\frac{g_{\sigma\sigma}}{V} \sum_{\k}\gamma_{\p \k},
   \\ \label{eq:equalb}
   \left(E_{\p\sigma;\p+\k\sigma;-\k}-  2\frac{g_{\sigma\sigma}}{V} \right) \rho_{\p \k} &=2\frac{g_{b \sigma}\sqrt{N}}{V}W_{\k}\alpha_{\p}+2\frac{g_{b \sigma}\sqrt{N}}{V}W_{\k}  \gamma_{\p, \k},
\\ \label{eq:equalc}
     \left(E_{\p+\k\sigma;\p-\k\sigma}-\frac{g_{\sigma\sigma}}{V}\right) \gamma_{\p \k} &=\frac{g_{b \sigma}\sqrt{N}}{V}W_{\k}\rho_{\p, -\k}+\frac{g_{\sigma\sigma}}{V}\alpha_{\p} +\frac{g_{\sigma\sigma}}{V}\sum_{\k'\neq \k} \gamma_{\p \k'}.
\end{align}
\end{subequations}
Here, we have used $E_{\p_1\sigma;\p_2\sigma'}=E-\epsilon_{\p_1\sigma}-\epsilon_{\p_2\sigma'}-g_{b\sigma}n-g_{b\sigma'}n$, and $E_{\p_1\sigma;\p_2\sigma';\k}=E-\epsilon_{\p_1\sigma}-\epsilon_{\p_2\sigma'}-E_{\k} -g_{b\sigma}n-g_{b\sigma'}n$.

We wish to calculate the energy to first order in the impurity-impurity interaction $g_{ \sigma\sigma}$ and second order in the impurity-medium one $g_{b \sigma}$. To do so, we first solve Eqs.~\eqref{eq:equalb} and \eqref{eq:equalc} to obtain $\rho_{\p \k}$ and $\gamma_{\p \k}$ at first and second order in $g_{b \sigma}$ respectively and then we insert the results in \eqref{eq:equala}. We obtain 
\begin{multline}
 E \simeq 2\left(\epsilon_{\p\sigma}+g_{b \sigma}n+\frac{g_{b \sigma}^2N}{V^2} \sum_{\k} \frac{W_\k^2}{ E_{\p\sigma;\p-\k\sigma;\k}-  2\frac{g_{\sigma\sigma}}{V} }\right)+\frac{g_{\sigma\sigma}}{V}
  \\ 
+\frac{4g_{\sigma\sigma}g_{b \sigma}^2n}{V^2} \sum_{\k} \frac{W_\k^2}{\left(E_{\p\sigma;\p-\k\sigma;\k}-  \frac{2g_{\sigma\sigma}}{V} \right) \left(E_{\p+\k\sigma;\p-\k\sigma}-\frac{g_{\sigma\sigma}}{V}\right)}.
\end{multline}
Inserting $E\simeq  2\epsilon_{\p \sigma}+2g_{b \sigma}n +\frac{g_{\sigma\sigma}}{V}$ in the denominators on the right hand side and expanding the first sum to first order in $g_{\sigma\sigma}$, we obtain the energy up to second order in $g_{b \sigma}$ 
\begin{multline}
 E \simeq 2\left(\epsilon_{\p_ \sigma}+g_{b \sigma}n+\frac{g_{b \sigma}^2n}{V} \sum_{\k} \frac{W_\k^2}{ \epsilon_{\p \sigma}-E_{\k}-\epsilon_{\p-\k \sigma}}\right)\\
\!+\!\frac{g_{\sigma\sigma}}{V}\left(\!1\! +\! 2\!  \frac{g_{b \sigma}^2n}{V} \sum_{\k} \left[\frac{2 W_\k^2}{ \left( \epsilon_{\p \sigma}-E_{\k}-\epsilon_{\p-\k \sigma} \right) \left( 2\epsilon_{\p \sigma}-\epsilon_{\p+\k \sigma}-\epsilon_{\p-\k \sigma}\right)}\!+\!\frac{W_\k^2}{ (\epsilon_{\p \sigma}-E_{\k}-\epsilon_{\p-\k \sigma})^2}\right]\right) .
\end{multline}
The energy has the form $E=2E_{\mathrm{pol}, \sigma}(\p)+\new{f_{\p\sigma,\p\sigma}}/V$, with the interaction
\begin{multline}
\label{eq:inter_boson_equalmom}
\new{f_{\p\sigma,\p\sigma}}
=g_{\sigma\sigma}\Bigg(1 + 2  \frac{g_{b \sigma}^2n}{V} \sum_{\k} \Bigg[\frac{2 W_\k^2}{ \left( \epsilon_{\p \sigma}-E_{\k}-\epsilon_{\p-\k \sigma} \right) \left( 2\epsilon_{\p \sigma}-\epsilon_{\p+\k \sigma}-\epsilon_{\p-\k \sigma}\right)}
\\
+\frac{W_\k^2}{ (\epsilon_{\p \sigma}-E_{\k}-\epsilon_{\p-\k \sigma})^2}\Bigg]\Bigg) .
\end{multline}

In the limit $\p\rightarrow 0$, this result was recently obtained in Ref.~\cite{levinsen2025}, in which case \new{we find the low-momentum quasiparticle interaction constant}
\begin{align}\label{eq:Fiipp}
F_{n,\sigma\sigma} 
& =g_{\sigma\sigma}\left( 1 + 2  \frac{g_{b \sigma}^2n}{V} \sum_{\k} \frac{W_\k^2}{E_{\k}+\epsilon_{\k \sigma} } \left[\frac{1}{  \epsilon_{\k \sigma}}+\frac{1}{E_{\k}+\epsilon_{\k \sigma}}\right]\right) .
\end{align}

\subsection{Distinguishable impurities}
For distinguishable impurities ($\sigma\neq \sigma'$), the ansatz instead takes the form
\begin{multline}
    \ket{\Psi}=\bigg[\alpha_{\p_1\p_2}c^\dag_{\p_1 \sigma}c^\dag_{\p_2 \sigma'}+\sum_{\k\neq0}\rho_{\p_1 \p_2 \k}c^\dag_{\p_1 \sigma}\beta^\dag_{-\k}c^\dag_{\p_2+\k \sigma'}+\sum_{\k\neq0}\eta_{\p_1 \p_2 \k}c^\dag _{\p_1+\k \sigma}\beta^\dag_{-\k}c^\dag_{\p_2 \sigma'}\\
    +\sum_\k \gamma_{\p_1 \p_2 \k}c^\dag _{\p_1+\k \sigma}c^\dag_{\p_2-\k \sigma'} \bigg]\ket{\Phi}. \label{eq:ansatzinter_dist}
\end{multline}
The equations of motion read
\begin{subequations} \label{eq:eqofmotinter_dist}
\begin{align} \nonumber
   \left(E_{\p_1 \sigma;\p_2 \sigma'}-\frac{g_{\sigma\sigma'}}{V}\right)\alpha_{\p_1\p_2} &=\frac{g_{b \sigma'}\sqrt{N}}{V} \sum_{\k} W_\k\rho_{\p_1 \p_2, -\k}  +\frac{g_{b \sigma}\sqrt{N}}{V} \sum_{\k} W_\k\eta_{\p_1 \p_2, -\k}  \\ \label{eq:eqofmotadist}
  & ~~ 
 +\frac{g_{\sigma\sigma'}}{V} \sum_{\k} \gamma_{\p_1 \p_2 \k}  ,
   \\ 
   \left(E_{\p_1 \sigma;\p_2+\k \sigma';-\k}-  \frac{g_{\sigma\sigma'}}{V}  \right) \rho_{\p_1 \p_2 \k} &= \frac{g_{b \sigma'}\sqrt{N}}{V}  W_\k\alpha_{\p_1\p_2} + \frac{g_{b \sigma}\sqrt{N}}{V} W_\k  \gamma_{\p_1 \p_2, -\k} +\frac{g_{\sigma\sigma'}}{V}  \eta_{\p_1 \p_2 \k} ,
   \\
     \left(E_{\p_1+\k \sigma;\p_2 \sigma';-\k}-\frac{g_{\sigma\sigma'}}{V}\right) \eta_{\p_1 \p_2 \k} &=  \frac{g_{b \sigma}\sqrt{N}}{V}  W_\k\alpha_{\p_1\p_2}+ \frac{g_{b \sigma'}\sqrt{N}}{V} W_\k   \gamma_{\p_1 \p_2 \k}
     +\frac{g_{\sigma\sigma'}}{V}   \rho_{\p_1 \p_2, \k}  ,
     \\ \nonumber
     \left(E_{\p_1+\k \sigma;\p_2-\k \sigma'}-\frac{g_{\sigma\sigma'}}{V}\right)\gamma_{\p_1 \p_2 \k} & =  \frac{g_{b \sigma}\sqrt{N}}{V}  W_\k\rho_{\p_1 \p_2, -\k}+ \frac{g_{b \sigma'}\sqrt{N}}{V}  W_{-\k}\eta_{\p_1 \p_2 \k}\\  & ~~+\frac{g_{\sigma\sigma'}}{V}  \alpha_{\p_1\p_2}  +\frac{g_{\sigma\sigma'}}{V}\sum_{\k'\neq\k}\gamma_{\p_1 \p_2 \k'}.
\end{align}
\end{subequations}
As before, we are interested in solving these equations perturbatively up to order second order in the impurity-medium interactions and leading order in the impurity-impurity one. To this end it is sufficient to evaluate $\rho_{\p_1 \p_2 \k},\eta_{\p_1 \p_2 \k}$  to first order and $\gamma_{\p_1 \p_2 \k}$ to second order and inject these in \eqref{eq:eqofmotadist}. Doing so, we obtain 
\begin{multline}
 E \simeq \epsilon_{\p_1 \sigma}+\epsilon_{\p_2 \sigma'}+g_{b \sigma}n+g_{b \sigma'}n+\frac{g_{\sigma\sigma'}}{V} 
 \\
 +\frac{g_{b \sigma'}^2 N}{V^2} \sum_{\k} \frac{W_\k^2}{  E_{\p_1 \sigma;\p_2-\k \sigma';\k}-  \frac{g_{\sigma\sigma'}}{V}  }
+\frac{g_{b \sigma}^2N}{V^2} \sum_{\k} \frac{ W_\k^2}{     E_{\p_1-\k \sigma;\p_2 \sigma';\k}- \frac{g_{\sigma\sigma'}}{V} }
\\ 
 + g_{\sigma\sigma'} \frac{g_{b \sigma} g_{b \sigma'}n}{V^2} \sum_{\k}\left[ \frac{2 W_\k^2}{ \left(E_{\p_1+\k \sigma;\p_2-\k \sigma'}-\frac{g_{\sigma\sigma'}}{V}\right) \left(  E_{\p_1 \sigma;\p_2-\k \sigma';\k}-  \frac{g_{\sigma\sigma'}}{V} \right)}\right.
 \\ 
 +\frac{ 2 W_\k^2}{ \left(E_{\p_1+\k \sigma;\p_2-\k \sigma'}-\frac{g_{\sigma\sigma'}}{V}\right)   \left(E_{\p_1+\k \sigma;\p_2 \sigma';\k}-  \frac{g_{\sigma\sigma'}}{V}\right) }
 \\ 
\left.+\frac{2 W_\k^2}{   \left(E_{\p_1-\k \sigma;\p_2 \sigma';\k}-  \frac{g_{\sigma\sigma'}}{V}\right) \left(  E_{\p_1 \sigma;\p_2-\k \sigma';\k}-  \frac{g_{\sigma\sigma'}}{V} \right)}\right].
\end{multline} 
Then, by inserting $E\simeq\epsilon_{\p_1 \sigma}+\epsilon_{\p_2 \sigma'}+g_{b \sigma}n+g_{b \sigma'}n+\frac{g_{\sigma\sigma'}}{V} $ on the right-hand side, we obtain an energy of the form $E=E_{\mathrm{pol} , \sigma}(\p_1)+E_{\mathrm{pol} , \sigma'}(\p_2)+\new{f_{\p_1\sigma,\p_2\sigma'}}
/V$, with a \new{quasiparticle} interaction of the form
\begin{multline}
\new{f_{\p_1\sigma,\p_2\sigma'}} 
=g_{\sigma\sigma'}\Biggl\{1
+\frac{g_{b \sigma} g_{b \sigma'}n}{V} \sum_{\k}\biggl[ \frac{ 2 W_\k^2}{ \left(\epsilon_{\p_1 \sigma}+\epsilon_{\p_2 \sigma'}-\epsilon_{\p_1+\k \sigma}-\epsilon_{\p_2-\k \sigma'}\right) \left(  \epsilon_{\p_2 \sigma'}-E_{\k}-\epsilon_{\p_2-\k \sigma'} \right)}
 \\ 
+\frac{2 W_\k^2}{ \left(\epsilon_{\p_1 \sigma}+\epsilon_{\p_2 \sigma'}-\epsilon_{\p_1+\k \sigma}-\epsilon_{\p_2-\k \sigma'}\right)  \left(\epsilon_{\p_1 \sigma}-E_{-\k}-\epsilon_{\p_1+\k \sigma}\right) }
 \\ 
+\frac{2 W_\k^2}{   \left(  \epsilon_{\p_2 \sigma'}-E_{\k}-\epsilon_{\p_2-\k \sigma'} \right) \left(\epsilon_{\p_1 \sigma}-E_\k-\epsilon_{\p_1-\k \sigma}\right) }\biggr]\Biggr\}. \label{eq:inter_dist}
\end{multline} 
We can observe that in the limit of $\p_1\rightarrow\p_2=\p$, Eq.~\eqref{eq:inter_dist} resembles the expression obtained above for the case of identical bosonic impurities at equal momenta \eqref{eq:inter_boson_equalmom}, and that it reduces to it if we take equal masses and interactions ($g_{b \sigma}=g_{b \sigma'}$ and $\epsilon_{\k \sigma}=\epsilon_{\k \sigma'}$). In the limit of vanishing momenta, \new{we obtain the quasiparticle interaction constant}
\begin{multline}
F_{n,\sigma\sigma'} 
=g_{\sigma\sigma'} \Biggl\{ 1 +\frac{ 2  g_{b\sigma} g_{b\sigma'}n}{V} \sum_{\k} \left[
 \frac{ W_\k^2}{ \epsilon_{\k \sigma}+\epsilon_{\k \sigma'}}\left(\frac{1}{  E_{\k}+\epsilon_{\k \sigma}}+\frac{1}{  E_{\k}+\epsilon_{\k \sigma'}}\right)\right.\\ \left.
 +  \frac{ W_\k^2}{   \left(  E_{\k}+\epsilon_{\k \sigma'} \right) \left(E_\k+\epsilon_{\k \sigma}\right)} \right] \Biggr\} .
\end{multline}

\subsection{Impurities with different momenta}
For impurities with \textit{different} momenta, the ansatz reads
\begin{multline}
\label{eq:ansatzinter_dist_ind_unequalp} 
    \ket{\Psi}=\Bigg[\alpha_{\p_1\p_2}c^\dag_{\p_1 \sigma}c^\dag_{\p_2 \sigma'}+\sum_{\k\neq0}\rho_{\p_1 \p_2 \k}c^\dag_{\p_1 \sigma}\beta^\dag_{-\k}c^\dag_{\p_2+\k \sigma'}+\sum_{\k\neq0}\eta_{\p_1 \p_2 \k}c^\dag _{\p_1+\k \sigma}\beta^\dag_{-\k}c^\dag_{\p_2 \sigma'} \\ 
+\sum_{\k\neq0}(1-\delta_{\sigma\sigma'}\delta_{\k,\p_2-\p_1})\gamma_{\p_1 \p_2 \k}c^\dag _{\p_1+\k \sigma}c^\dag_{\p_2-\k \sigma'} \Bigg]\ket{\Phi} 
\end{multline}
We note that for indistinguishable impurities, the last sum must exclude $\k=\p_2-\p_1$ since otherwise we simply reproduce the first term. Furthermore, we have $\gamma_{\p_1 \p_2, \p_2-\p_1-\k}=\gamma_{\p_1 \p_2, \k}$, and $\gamma_{\p_1 \p_2, \p_1-\p_2+\k}=\gamma_{\p_1 \p_2, -\k}$ (these relations generalize  the relation $\gamma_{\p, -\k}=\gamma_{\p, \k}$ for equal momentum impurities).

The equations of motion read
\begin{subequations} \label{eq:EOMinter_A3}
\begin{align} \nonumber
   \left(E_{\p_1 \sigma;\p_2 \sigma'}-\frac{g_{\sigma\sigma'}}{V} (1\pm\delta_{\sigma\sigma'})\right)\alpha_{\p_1\p_2} & = 
  \frac{g_{b \sigma'}\sqrt{N}}{V} \sum_{\k} W_\k\rho_{\p_1 \p_2, -\k}+\frac{g_{b \sigma}\sqrt{N}}{V} \sum_{\k} W_\k\eta_{\p_1 \p_2, -\k} \\ \nonumber
   & +  \frac{g_{\sigma\sigma'}}{V} (1\pm\delta_{\sigma\sigma'})\sum_{\k} \gamma_{\p_1 \p_2 \k} (1-\delta_{\sigma\sigma'}\delta_{\k,\p_2-\p_1})
  \\ \label{eq:eqofmotintera}
  &  \hspace{-20mm}    \pm \delta_{\sigma\sigma'} \frac{g_{b \sigma'}\sqrt{N}}{V} \left(W_{\p_2-\p_1}\rho_{\p_1 \p_2, \p_1-\p_2}+W_{\p_1-\p_2}\eta_{\p_1 \p_2, \p_2-\p_1}\right) 
   ,
   \\\nonumber
    \left(E_{\p_1 \sigma;\p_2+\k \sigma',-\k}-  \frac{g_{\sigma\sigma'}}{V} (1\pm\delta_{\sigma\sigma'})  \right) \rho_{\p_1 \p_2 \k}   &=   \frac{g_{b \sigma'}\sqrt{N}}{V}  W_\k\alpha_{\p_1\p_2}  +\frac{g_{b \sigma}\sqrt{N}}{V} W_\k  \gamma_{\p_1 \p_2, -\k}(1\pm\delta_{\sigma\sigma'})
    \\
    & +\frac{g_{\sigma\sigma'}}{V} (1\pm\delta_{\sigma\sigma'}) \eta_{\p_1 \p_2 \k} , \label{eq:eqofmotinterb}
\end{align}
\end{subequations}
\setcounter{equation}{\value{equation}-1}
\begin{subequations}
\setcounter{equation}{2}
\begin{align}
\nonumber
   \left(E_{\p_1 \sigma;\p_1 \sigma';\p_2-\p_1}-  \frac{g_{\sigma\sigma'}}{V}   \right) \rho_{\p_1 \p_2, \p_1-\p_2} 
   & =   \frac{g_{b \sigma'}\sqrt{N}}{V}  W_{\p_1-\p_2}\alpha_{\p_1\p_2} 
   \\ \nonumber
   &    +\frac{g_{b \sigma}\sqrt{N}}{V}  W_{\p_1-\p_2} \gamma_{\p_1 \p_2, \p_2-\p_1}
    (1-\delta_{\sigma\sigma'})
    \\ 
   &+\frac{g_{\sigma\sigma'}}{V}\eta_{\p_1 \p_2, \p_1-\p_2}  
   , \label{eq:eqofmotinterb2}
   \\ \nonumber
       \left(E_{\p_1+\k \sigma;\p_2 \sigma';-\k}- \frac{g_{\sigma\sigma'}}{V}(1\pm\delta_{\sigma\sigma'})\right) \eta_{\p_1 \p_2 \k}    & = 
     \frac{g_{b \sigma}\sqrt{N}}{V}  W_\k\alpha_{\p_1\p_2}+ \frac{g_{b \sigma'}\sqrt{N}}{V} W_{\k}  \gamma_{\p_1 \p_2 \k}(1\pm\delta_{\sigma\sigma'})\\
     &+\frac{g_{\sigma\sigma'}}{V} (1\pm\delta_{\sigma\sigma'})  \rho_{\p_1 \p_2 \k} , \label{eq:eqofmotinterc}
        \\  \nonumber
       \left(E_{\p_2 \sigma;\p_2 \sigma';\p_1-\p_2}-  \frac{g_{\sigma\sigma'}}{V}\right) \eta_{\p_1 \p_2, \p_2-\p_1}  
     & =  \frac{g_{b \sigma}\sqrt{N}}{V}  W_{\p_2-\p_1}\alpha_{\p_1\p_2}  
      \\ \nonumber
      &   + \frac{g_{b \sigma'}\sqrt{N}}{V} W_{\p_2-\p_1}  \gamma_{\p_1 \p_2, \p_2-\p_1} (1-\delta_{\sigma\sigma'})
      \\
      & 
      +\frac{g_{\sigma\sigma'}}{V}  \rho_{\p_1 \p_2, \p_2-\p_1}  
      , \label{eq:eqofmotinterc2}
    \\ \nonumber
     \left(E_{\p_1+\k \sigma;\p_2-\k \sigma'}-\frac{g_{\sigma\sigma'}}{V}\right) \gamma_{\p_1 \p_2 \k}( 1\pm \delta_{\sigma\sigma'} )&=  \nonumber
     \frac{g_{b \sigma}\sqrt{N}}{V}  \left(W_\k\rho_{\p_1 \p_2, -\k}\pm\delta_{\sigma\sigma'} W_{\p_2-\p_1-\k}\rho_{\p_1 \p_2, \p_1-\p_2+\k}\right) \\  \nonumber
   & +\frac{g_{b \sigma'}\sqrt{N}}{V}  \left(W_{-\k}\eta_{\p_1 \p_2 \k}\pm\delta_{\sigma\sigma'} W_{\p_1-\p_2+\k}\eta_{\p_1 \p_2, \p_2-\p_1-\k}\right)
   \\  \nonumber
   &    +\frac{g_{\sigma\sigma'}}{V} (1\pm\delta_{\sigma\sigma'}) \sum_{\k'\neq\k}\gamma_{\p_1 \p_2 \k'} (1-\delta_{\sigma\sigma'}\delta_{\k',\p_2-\p_1})\\
    &   +\frac{g_{\sigma\sigma'}}{V} (1\pm\delta_{\sigma\sigma'}) \alpha_{\p_1\p_2}  \,.
    \label{eq:eqofmotinterd}
\end{align}
\end{subequations}
We recall that the top sign in the $\pm\delta_{\sigma\sigma'}$ terms corresponds to bosonic impurities while the bottom sign corresponds to fermionic impurities. Here, we have explicitly separated the cases $\k=\p_1-\p_2$ and  $\k=\p_2-\p_1$ in Eqs.~\eqref{eq:eqofmotinterb2} and \eqref{eq:eqofmotinterc2}. We remark that for identical bosonic impurities, Eq.~\eqref{eq:eqofmotintera} contains an extra direct interaction terms $\frac{g_{\sigma\sigma}}{V}$ compared to the case $\p_1=\p_2$ presented in the previous subsection [Eq. \eqref{eq:equala}]. In the equal-momentum case, this term is absent because it cancels with the factor two originating from $   \bra{\Phi}c_{\p_2 \sigma'}c_{\p_1 \sigma}\ket{\Psi} =\alpha_{\p_1\p_2}(1+\delta_{\sigma\sigma'}\delta_{\p_1\p_2})$.

As before, we are interested in solving these equations perturbatively up to second order in the impurity-medium interactions and first order in the impurity-impurity interactions. It is thus sufficient to evaluate $\rho_{\p_1 \p_2 \k},\eta_{\p_1 \p_2 \k}$ to first order and $\gamma_{\p_1 \p_2 \k}$ to second order in the impurity-medium interaction respectively and insert the results in \eqref{eq:eqofmotintera}. Doing so, we obtain
\begin{multline} 
 E\simeq \epsilon_{\p_1 \sigma}+\epsilon_{\p_2 \sigma'}+g_{b \sigma}n+g_{b \sigma'}n+\frac{g_{\sigma\sigma'}}{V}(1\pm\delta_{\sigma\sigma'}) \\
 +\frac{g_{b \sigma'}^2 N}{V^2} \sum_{\k} \frac{W_\k^2}{  E_{\p_1\sigma;\p_2-\k\sigma';\k}-\frac{g_{\sigma\sigma'}(1\pm\delta_{\sigma\sigma'})}{V}}
 +\frac{g_{b \sigma}^2N}{V^2} \sum_{\k} \frac{W_\k^2}{   E_{\p_1-\k\sigma;\p_2\sigma';\k}-\frac{g_{\sigma\sigma'}(1\pm\delta_{\sigma\sigma'})}{V}}
   \\  \pm\delta_{\sigma\sigma'} \frac{g_{b \sigma}^2N}{V^2} \left[  \frac{W_{\p_2-\p_1}^2}{  E_{\p_1\sigma;\p_1\sigma';\p_2-\p_1}-\frac{g_{\sigma\sigma'}(1\pm\delta_{\sigma\sigma'})}{V}}
 + \frac{W_{\p_1-\p_2}^2}{E_{\p_2\sigma;\p_2\sigma';\p_1-\p_2}-\frac{g_{\sigma\sigma'}(1\pm\delta_{\sigma\sigma'})}{V}}\right]
 \\  + \frac{g_{\sigma\sigma'} }{V} \frac{g_{b \sigma} g_{b \sigma'}N}{V^2} \sum_{\k} \Biggl[  \frac{2 W_\k^2  (1\pm\delta_{\sigma\sigma'}) }{ \left(E_{\p_1\sigma;\p_2-\k\sigma';\k}-\frac{g_{\sigma\sigma'}(1\pm\delta_{\sigma\sigma'})}{V}\right)\left(  E_{\p_1-\k\sigma;\p_2\sigma';\k}-\frac{g_{\sigma\sigma'}(1\pm\delta_{\sigma\sigma'})}{V}\right)}
\\   +
\frac{ W_\k^2  [2-\delta_{\k,\p_2-\p_1}(1+\delta_{\sigma\sigma'})]  (1\pm\delta_{\sigma\sigma'})}{ \left(E_{\p_1\sigma;\p_2-\k\sigma';\k}-\frac{g_{\sigma\sigma'}(1\pm\delta_{\sigma\sigma'})}{V}\right) \left(E_{\p_1+\k\sigma;\p_2-\k\sigma'}-\frac{g_{\sigma\sigma'}}{V}\right)}
 \\ + 
 \frac{ W_\k^2    [2-\delta_{\k,\p_2-\p_1}(1+\delta_{\sigma\sigma'})]  (1\pm\delta_{\sigma\sigma'}) }{\left(E_{\p_1+\k\sigma;\p_2\sigma';-\k}-\frac{g_{\sigma\sigma'}(1\pm\delta_{\sigma\sigma'})}{V}\right) \left(E_{\p_1+\k\sigma;\p_2-\k\sigma'}-\frac{g_{\sigma\sigma'}}{V}\right)}\Biggr]
\\  
  + \frac{g_{\sigma\sigma'} }{V} \frac{g_{b \sigma} g_{b \sigma'}N}{V^2}  \Bigg[ \frac{ W_{\p_2-\p_1}^2 }{ \left(E_{\p_1\sigma;\p_1\sigma';\p_2-\p_1}-\frac{g_{\sigma\sigma'}(1\pm\delta_{\sigma\sigma'})}{V}\right) \left(E_{\p_2\sigma;\p_1\sigma'}-\frac{g_{\sigma\sigma'}}{V}\right)} 
  \\
 +\frac{ W_{\p_1-\p_2}^2 }{\left(E_{\p_2\sigma;\p_2\sigma';\p_1-\p_2}-\frac{g_{\sigma\sigma'}(1\pm\delta_{\sigma\sigma'})}{V}\right) \left(E_{\p_2\sigma;\p_1\sigma'}-\frac{g_{\sigma\sigma'}}{V}\right)}\Bigg](1-\delta_{\sigma\sigma'}),
\end{multline}
Then, by inserting $E\simeq\epsilon_{\p_1 \sigma}+\epsilon_{\p_2 \sigma'}+g_{b \sigma}n+g_{b \sigma'}n+\frac{g_{\sigma\sigma'}}{V} (1\pm\delta_{\sigma\sigma'}) $ in the different denominators on the right-hand side, we obtain the energy up to second order in $g_{b \sigma}$. As above, we find that it takes the form $E=E_{\mathrm{pol}, \sigma}(\p_1)+E_{\mathrm{pol} ,\sigma'}(\p_2)+
\new{f_{\p_1\sigma,\p_2\sigma'}}
/V$, with the polaron-polaron interactions
\begin{multline}
\new{f_{\p_1\sigma,\p_2\sigma'}}
= \pm \delta_{\sigma\sigma'}g_{ b  \sigma}^2nW_{\p_2-\p_1}^2 \left(\frac{1}{\epsilon_{\p_1 \sigma}-\epsilon_{\p_2 \sigma}-E_{\p_2-\p_1}}+\frac{1}{ \epsilon_{\p_2 \sigma}-\epsilon_{\p_1 \sigma}-E_{\p_2-\p_1}} \right) 
\\ 
+  (1\pm\delta_{\sigma\sigma'})  g_{\sigma\sigma'}\Biggl\{1+
\frac{g_{b \sigma} g_{b \sigma'}n}{V} \sum_{\k} \left[\frac{2 W_\k^2}{   \left(  \epsilon_{\p_2 \sigma'}-E_{-\k}-\epsilon_{\p_2+\k \sigma'} \right) \left(\epsilon_{\p_1 \sigma}-E_{-\k}-\epsilon_{\p_1+\k \sigma}\right) }\right.
\\ 
+\frac{2  W_\k^2  }{ \left(\epsilon_{\p_1 \sigma}+\epsilon_{\p_2 \sigma'}-\epsilon_{\p_1+\k \sigma}-\epsilon_{\p_2-\k \sigma'}\right) \left(  \epsilon_{\p_2 \sigma'}-E_{\k}-\epsilon_{\p_2-\k \sigma'} \right)}
 \\ 
\left.+\frac{2 W_\k^2 }{ \left(\epsilon_{\p_1 \sigma}+\epsilon_{\p_2 \sigma'}-\epsilon_{\p_1+\k \sigma}-\epsilon_{\p_2-\k \sigma'}\right)  \left(\epsilon_{\p_1 \sigma}-E_{-\k}-\epsilon_{\p_1+\k \sigma}\right) }\right]\Biggr\}. \label{eq:higher1/V_removed}
\end{multline} 
Here we have neglected terms which are of higher order in $1/V$. The term in the first line is independent of the bare impurity-impurity interactions, and corresponds to the induced interactions for non-interacting impurities derived in the main text. The other terms only exist if $g_{\sigma\sigma'}\neq0$.  

First, we can observe that this latter part reduces to the result obtained in Eq.~\eqref{eq:inter_dist} for distinguishable impurities when $ \delta_{\sigma\sigma'}=0$, as it should. On the other hand, for indistinguishable impurities ($\delta_{\sigma\sigma'}=1$), we can see that the bare impurity-impurity interaction only plays a role when these are bosons. In this case, taking the limit $\p_1,\p_2\rightarrow 0$, the polaron interactions take the form
\begin{align} \label{eq:ppinter_indis}
F_{n,\sigma\sigma}
& = \begin{cases}
    - \frac{g_{ b  \sigma}^2}{g_{bb}}
+ 2 g_{ \sigma \sigma}\left[1+2\frac{g_{b \sigma}^2n}{V} \sum_{\k} \frac{ W_\k^2}{ E_{\k}+\epsilon_{\k \sigma} }  \left( \frac{1}{ E_{\k}+\epsilon_{\k \sigma} } +\frac{ 1 
}{\epsilon_{\k \sigma}}\right)\right] & \text{bosons} \\
+\frac{g_{ b  \sigma}^2}{g_{bb}}&  \text{fermions} 
  \end{cases}\,.
\end{align}
We see that the contribution due to the bare interaction in \eqref{eq:ppinter_indis} resembles the one obtained for degenerate bosonic impurities taking $\p_1=\p_2=0$ from the start, Eq.~\eqref{eq:Fiipp}, but with an overall factor of 2 due to the possibility of exchanging bosons with distinct momenta. 

\section{Interacting Fermi polarons}
\setcounter{equation}{0}
\label{app:intFermi}
We provide in this appendix the details of the derivation of the Fermi-polaron interaction in Eq.~\eqref{eq:inducedinterfermioninterimp} for the case where the bare impurity interaction strength $g_{\sigma \sigma'}$ is finite. Our starting point is the variational ansatz~\eqref{eq:2-pol-F-med-int}. We remind the reader that we assume $\p_1\ne \p_2$ for identical ($\sigma=\sigma'$) fermionic impurities, and we take $k>k_F\geq q$ throughout this appendix. Also, for indistinguishable bosonic impurities, $\gamma_{\p_1\p_2\K}$ must satisfy the symmetry conditions $\gamma_{\p_1\p_2, \p_2 - \p_1 -\K} = \gamma_{\p_1\p_2, \K}$.

The normalization of the state~\eqref{eq:2-pol-F-med-int} depends on whether we are considering indistinguishable bosonic impurities with equal or different momenta, indistinguishable fermionic impurities (with $\p_1 \ne \p_2$), or distinguishable impurities:
\begin{multline}
\label{eq:2-pol-F-med-nor-int}
    1 = \bra{\Psi} \ket{\Psi} = (1+\delta_{\p_1\p_2} \delta_{\sigma\sigma'}) |\alpha_{\p_1\p_2}|^2 + \sum_{\k\q} |\rho_{\p_1\p_2\k\q}|^2 + (1-\delta_{\p_1\p_2} \delta_{\sigma\sigma'}) \sum_{\k\q} |\eta_{\p_1\p_2\k\q}|^2 \\
    \pm \delta_{\sigma\sigma'} (1-\delta_{\p_1\p_2}) \left[\sum_{\k\q} \left(\delta_{\q-\k , \p_1-\p_2}|\rho_{\p_1\p_2\k\q}|^2 + \delta_{\q-\k , \p_2-\p_1}|\eta_{\p_1\p_2\k\q}|^2\right)\right]\\
    + \sum_{\K\neq0}|\gamma_{\p_1\p_2\K}|^2\left(1\pm \delta_{\sigma\sigma'}\right) \left(1- \delta_{\sigma\sigma'}\delta_{\K,\p_2-\p_1}\right)\; .
\end{multline}
For clarity, let's consider the three cases separately.

\subsection{Indistinguishable bosonic impurities with equal momenta}
\begin{figure}[tbp]    
\centering
\includegraphics[width=.5\linewidth]{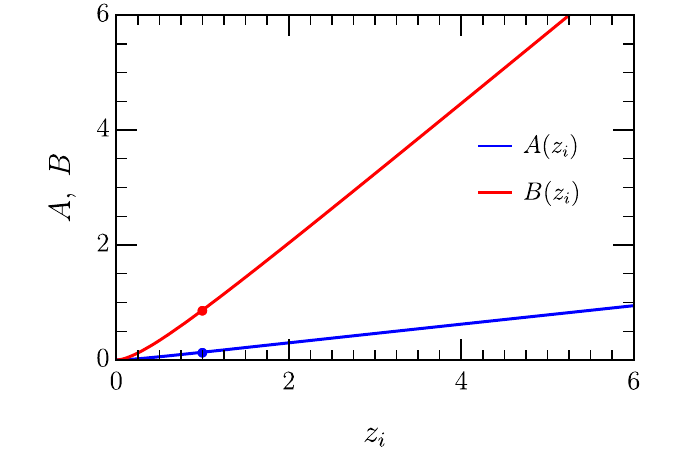}
\caption{Mass ratio $z_\sigma =m_\sigma /m_{b,f}$ dependence of the functions $A(z_\sigma )$~\eqref{eq:A-fun} and $B(z_\sigma )$~\eqref{eq:B-func} appearing in the expressions of the Bose (blue line) and Fermi (red line) polaron interactions. The dots indicate the values for equal masses with $z_\sigma =1$ --- $A(1)=4/3\pi^2$ and $B(1)=1/4 + \pi^2/16$.}
\label{fig:AB_functions}
\end{figure}

In this case, there is no contribution from the $\eta_{\p \p \k\q}$ term in Eq.~\eqref{eq:2-pol-F-med-int} and in the normalization~\eqref{eq:2-pol-F-med-nor-int}. The variational two-polaron state and its normalization are:
\begin{subequations}
\begin{align}
  \ket{\Psi} &= \left( \alpha_{\p} c^{\dag}_{\p \sigma} c^{\dag}_{\p \sigma} +
  \sum_{\k\q} \rho_{\p \k\q}
  c^{\dag}_{\p \sigma} f^{\dag}_{\k} f_{\q}^{} c^{\dag}_{\p -\k+\q \sigma}  +  \sum_{\k \ne \0} \gamma_{\p \K} c^{\dag}_{\p + \K \sigma} c^{\dag}_{\p -\K \sigma} \right) \fs\\
  1 &= \bra{\Psi} \ket{\Psi} = 2 |\alpha_{\p}|^2 + \sum_{\k\q} |\rho_{\p \k\q}|^2 + 2 \sum_{\K \ne \0} |\gamma_{\p \K}|^2 \; .
\end{align}
\end{subequations}
The equations of motion are given by:
\begin{subequations}
\begin{align}
    E \alpha_{\p} &= \left(2 \epsilon_{\p \sigma} + 2 g_{f\sigma} n + \frac{g_{\sigma\sigma}}{V} \right)\alpha_{\p} + \frac{g_{f\sigma}}{V} \sum_{\k\q} \rho_{\p\k\q} + \frac{g_{\sigma\sigma}}{V} \sum_{\K\ne \0} \gamma_{ \p \K} \label{eq:alpha-same-mom}\\
    E \rho_{\p \k\q} &= \left(\epsilon_{\p \sigma} + E_{\p\k\q} + 2 g_{f\sigma} n + 2 \frac{g_{\sigma\sigma}}{V} \right)\rho_{\p \k\q}   + 2 \frac{g_{f\sigma}}{V} \alpha_{\p} +  2 \frac{g_{f\sigma}}{V} \gamma_{\p, \k -\q } \label{eq:rho-same-mom}\\
    E \gamma_{\p \K } &= \left(\epsilon_{\p + \K \sigma} + \epsilon_{\p - \K \sigma} + 2 g_{f\sigma} n + \frac{g_{\sigma\sigma}}{V}\right)\gamma_{ \p \K} + \frac{g_{\sigma\sigma}}{V} \alpha_\p + \frac{g_{f\sigma}}{V} \sum_{\q'} \rho_{\p, \q' -\K, \q'} \; .
    \label{eq:gamma-same-mom}
\end{align}
\end{subequations}
Because we look for the perturbative corrections of the two polaron energy expression up to second order in $g_{f\sigma}$ and first order in $g_{\sigma\sigma}$, we can solve the equation for $\rho_{\p \k\q}$~\eqref{eq:rho-same-mom} up to first order in both $g_{f\sigma}$ and $g_{\sigma\sigma}$, and the equation for $\gamma_{\p \K }$~\eqref{eq:gamma-same-mom} to second order in $g_{f\sigma}$ and zeroth order in $g_{\sigma\sigma}$. By substituting these solutions back into Eq.~\eqref{eq:alpha-same-mom}, we obtain:
\begin{multline}
    E \simeq  2 E_{\mathrm{pol},\sigma} (\p) + \frac{g_{\sigma\sigma}}{V}\\
    + 2\frac{g_{f\sigma}^2 g_{\sigma\sigma}}{V^3} \sum_{\k \q}\frac{1}{\epsilon_{\p \sigma} - E_{\p\k\q \sigma}}\left[\frac{1}{\epsilon_{\p \sigma} - E_{\p \k\q \sigma}} + \frac{2}{2\epsilon_{\p \sigma} - \epsilon_{\p+\k-\q \sigma} - \epsilon_{\p-\k+\q \sigma}}\right] \; .
\label{eq:FP_ind-imp_same-momenta}
\end{multline}
Thus, in the limit $\p\to\0$, the polaron interaction takes the form
\begin{align}
\label{eq:Fermi-pol-int_same-mom}
F_{n,\sigma\sigma}
& \simeq g_{\sigma\sigma}\left[1 + 2\frac{g_{f\sigma}^2}{V^2} \sum_{\k \q}\frac{1}{\epsilon_{\q-\k \sigma} + \epsilon_\k-\epsilon_\q}\left(\frac{1}{\epsilon_{\q-\k \sigma} + \epsilon_\k-\epsilon_\q} + \frac{1}{\epsilon_{\q-\k \sigma}}\right)\right]\\
 &   = g_{\sigma\sigma}\left[1 + 9 \left(\frac{g_{f\sigma} n}{E_F}\right)^2B(z_\sigma )\right]\; ,
\end{align}
where $z_\sigma  = m_\sigma /m_f$ is the mass ratio. 
The function $B(z_\sigma )$, where
\begin{subequations}
\label{eq:B-func}
\begin{align}
\label{eq:B-fun}
    B(z_\sigma ) &= f_{1}(z_\sigma ) + f_2(z_\sigma )\\
    f_1(z_\sigma ) &= \int_{1}^\infty dk k^2 \int_{-1}^1 dx \int_{0}^1 dq q^2 \left[\frac{q^2 + k^2 -2 q k x}{z_\sigma} +k^2 -q^2\right]^{-2}\nonumber\\
    &= \frac{z_\sigma ^2 \log z_\sigma}{2
   (z_\sigma ^2-1)} \To_{z_\sigma \to 1} \frac{1}{4}\\
    f_2(z_\sigma ) &=\int_{1}^\infty dk k^2 \int_{-1}^1 dx \int_{0}^1 dq q^2 \left[\frac{q^2 + k^2 -2 q k x}{z_\sigma}\left(\frac{q^2 + k^2 -2 q k x}{z_\sigma} +k^2 -q^2\right)\right]^{-1}\nonumber\\
    &\To_{z_\sigma \to 1} \frac{\pi^2}{16}\; ,
\end{align}
\end{subequations}
is plotted in Fig.~\ref{fig:AB_functions}.

\subsection{Indistinguishable bosonic impurities with different momenta}
We start from the following variational state with $\p_1 \ne \p_2$:
\begin{multline}
    \ket{\Psi} = \left[ \alpha_{\p_1\p_2} c^{\dag}_{\p_1 \sigma} c^{\dag}_{\p_2 \sigma} +
  \sum_{\k\q} \rho_{\p_1\p_2\k\q}
  c^{\dag}_{\p_1\sigma} f^{\dag}_{\k} f_{\q}^{} c^{\dag}_{\p_2 -\k+\q \sigma}\right.\\
  \left.  + \sum_{\k\q} \eta_{\p_1\p_2\k\q}
  c^{\dag}_{\p_1 -\k+\q \sigma} f^{\dag}_{\k} f_{\q}^{} c^{\dag}_{\p_2 \sigma} + \sum_{\K\neq0}(1-\delta_{\K,\p_2-\p_1})\gamma_{\p_1\p_2\K}c^\dag _{\p_1+\K \sigma}c^\dag_{\p_2-\K \sigma}  \right]\fs\; .
\end{multline}
Note that the term $\sum_{\k\q} \rho_{\p_1\p_2\k\q}$ ($\sum_{\k\q} \eta_{\p_1\p_2\k\q}$) contains also the terms with $\q-\k = \p_1-\p_2$ ($\q-\k = \p_2-\p_1$) for which the two impurities have both momenta equal to $\p_1$ ($\p_2$). Thus, the normalization is:
\begin{multline}
\label{eq:2-pol-F-med-nor-int2}
    1 = \bra{\Psi} \ket{\Psi} =  |\alpha_{\p_1\p_2}|^2 + \sum_{\k\q} |\rho_{\p_1\p_2\k\q}|^2 +  \sum_{\k\q} |\eta_{\p_1\p_2\k\q}|^2\\
      + \left[\sum_{\k\q} \left(\delta_{\q-\k , \p_1-\p_2}|\rho_{\p_1\p_2\k\q}|^2 + \delta_{\q-\k , \p_2-\p_1}|\eta_{\p_1\p_2\k\q}|^2\right)\right] + 2 \sum_{\K\neq0}|\gamma_{\p_1\p_2\K}|^2 \left(1- \delta_{\K,\p_2-\p_1}\right)\; ,
\end{multline}
where we have made use of the symmetry $\gamma_{\p_1\p_2, \p_2 - \p_1 -\K} = \gamma_{\p_1\p_2, \K}$. 
Because of their different normalization, we separate the equations of motion for the $\rho_{\p_1\p_2\k\q}$ ($\eta_{\p_1\p_2\k\q}$) terms with $\q-\k\ne \p_1-\p_2$ ($\q-\k \ne \p_2 - \p_1$) and for $\rho_{\p_1\p_2, \k, \k+ \p_1 - \p_2}$ ($\eta_{\p_1\p_2, \k, \k+ \p_2 - \p_1}$), obtaining
\begin{subequations}\label{eq:longandcomplicated}
\begin{align}
    E \alpha_{\p_1\p_2} =& \left(\epsilon_{\p_1 \sigma} + \epsilon_{\p_2 \sigma} + 2 g_{f\sigma} n + 2 \frac{g_{\sigma\sigma}}{V} \right)\alpha_{\p_1\p_2}\nonumber\\ 
    &+ \frac{g_{f\sigma}}{V} \sum_{\k\q} (1-\delta_{\q-\k,\p_1 - \p_2}) \rho_{\p_1\p_2, \k\q} +2
    \frac{g_{f\sigma}}{V} \sum_{\k} \rho_{\p_1\p_2, \k, \k+ \p_1 - \p_2} \nonumber\\
    & + \frac{g_{f\sigma}}{V} \sum_{\k\q} (1-\delta_{\q-\k,\p_2 - \p_1}) \eta_{\p_1\p_2, \k\q} +  2 \frac{g_{f\sigma}}{V} \sum_{\k} \eta_{\p_1\p_2, \k, \k+ \p_2 - \p_1}\nonumber\\ 
    &+ 2 \frac{g_{\sigma\sigma}}{V} \sum_{\K} (1-\delta_{\K,\p_2-\p_1}) \gamma_{ \p_1 \p_2\K} \label{eq:aint}\\
    E \rho_{\p_1\p_2, \k\q} =& \left(\epsilon_{\p_1 \sigma} + E_{\p_2,\k\q \sigma} + 2g_{f\sigma} n + 2 \frac{g_{\sigma\sigma}}{V} \right)\rho_{\p_1\p_2, \k\q}   + \frac{g_{f\sigma}}{V} \alpha_{\p_1\p_2} \nonumber \\   
    &+  2 \frac{g_{f\sigma}}{V} \gamma_{\p_1 \p_2, \k -\q } + 2 \frac{g_{\sigma\sigma}}{V} \eta_{\p_1\p_2, \k\q}
      \label{eq:b_in}\\
    2 E \rho_{\p_1\p_2, \k, \k+ \p_1 - \p_2} =& 2 \left(2 \epsilon_{\p_1 \sigma} + \epsilon_{\k} - \epsilon_{\p_1 - \p_2 + \k} + 2g_{f\sigma} n  + \frac{g_{\sigma\sigma}}{V} \right)\rho_{\p_1\p_2, \k, \k+ \p_1 - \p_2}\nonumber \\   
    &+ 2 \frac{g_{f\sigma}}{V} \alpha_{\p_1\p_2} + \frac{2 g_{\sigma\sigma}}{V} \eta_{\p_1\p_2, \k,\k+ \p_1-\p_2}\label{eq:cint}\\
    E \eta_{\p_1\p_2, \k\q} =& \left(\epsilon_{\p_2 \sigma} + E_{\p_1,\k\q \sigma} + 2 g_{f\sigma} n + 2 \frac{g_{\sigma\sigma}}{V}\right)\eta_{\p_1\p_2, \k\q} + \frac{g_{f\sigma}}{V} \alpha_{\p_1\p_2} \nonumber \\   
    &+ 2 \frac{g_{f\sigma}}{V} \gamma_{\p_1 \p_2, \q -\k } + 2 \frac{g_{\sigma\sigma}}{V} \rho_{\p_1\p_2, \k\q}\label{eq:dint}
        \\
    2E \eta_{\p_1\p_2, \k, \k+ \p_2 - \p_1} =& 2 \left(2 \epsilon_{\p_2 \sigma} + \epsilon_{\k} - \epsilon_{\p_2 - \p_1 + \k} + 2g_{f\sigma} n  + \frac{g_{\sigma\sigma}}{V}  \right)\eta_{\p_1\p_2, \k, \k+ \p_2 - \p_1} \nonumber\\
     &+ 2 \frac{g_{f\sigma}}{V} \alpha_{\p_1\p_2} + 2 \frac{g_{\sigma\sigma}}{V} \rho_{\p_1\p_2, \k,\p_2-\p_1+\k}\label{eq:eint} 
\end{align}
\end{subequations}
\setcounter{equation}{\value{equation}-1}
\begin{subequations}
\setcounter{equation}{5}
\begin{align}
    2 E \gamma_{\p_1 \p_2, \K } =& 2 \left(\epsilon_{\p_1 + \K \sigma} + \epsilon_{\p_2 - \K \sigma} + 2 g_{f\sigma} n + \frac{g_{\sigma\sigma}}{V}\right)\gamma_{ \p_1 \p_2, \K} \nonumber\\
     &+ 2 \frac{g_{\sigma\sigma}}{V} \sum_{\K'\ne\K}\gamma_{\p_1 \p_2, \K'}  + 2 \frac{g_{\sigma\sigma}}{V} \alpha_{\p_1 \p_2} \nonumber\\
    & + \frac{g_{f\sigma}}{V} \sum_{\k'\q'} (1-\delta_{\q'-\k',\p_1 - \p_2}) \rho_{\p_1 \p_2, \k'\q'} \left(\delta_{\k'-\q',\K}  + \delta_{\q'-\k',\K + \p_1 - \p_2} \right) \nonumber\\
    & + \frac{g_{f\sigma}}{V} \sum_{\k'\q'} (1-\delta_{\q'-\k',\p_2 - \p_1}) \eta_{\p_1 \p_2, \k'\q'} \left(\delta_{\q'-\k',\K} + \delta_{\k'-\q',\K + \p_1 - \p_2} \right) \; .
\end{align}
\end{subequations}

The strategy to solve these equations perturbatively, up to first order in $g_{\sigma\sigma}$ and second order in $g_{f\sigma}$, goes exactly as described before, giving
$E = E_{\mathrm{pol},\sigma} (\p_1) + E_{\mathrm{pol},\sigma} (\p_2) +
\new{f_{\p_1\sigma,\p_2\sigma'}}/V$, where
\begin{multline}
    \new{f_{\p_1\sigma,\p_2\sigma'}}
    \simeq  2 g_{\sigma\sigma} + \frac{2 g_{f\sigma}^2 g_{\sigma\sigma}}{V^2} \sum_{\k\q} \frac{\frac{1}{\epsilon_{\p_1 \sigma} + \epsilon_{\p_2 \sigma} -\epsilon_{\p_1+\k-\q \sigma} - \epsilon_{\p_2-\k+\q \sigma}} + \frac{1}{\epsilon_{\p_1 \sigma} - \epsilon_{\p_1+\q-\k \sigma} -\epsilon_\k+\epsilon_\q}}{\epsilon_{\p_2 \sigma} - \epsilon_{\p_2+\q-\k \sigma} -\epsilon_\k+\epsilon_\q}
    \\
    + \frac{2g_{f\sigma}^2 g_{\sigma\sigma}}{V^2} \sum_{\k\q} \frac{\frac{1}{\epsilon_{\p_1 \sigma} + \epsilon_{\p_2 \sigma} -\epsilon_{\p_2+\k-\q \sigma} - \epsilon_{\p_1-\k+\q \sigma}} + \frac{1}{\epsilon_{\p_2 \sigma} - \epsilon_{\p_2+\q-\k \sigma} -\epsilon_\k+\epsilon_\q}}{\epsilon_{\p_1 \sigma} - \epsilon_{\p_1+\q-\k \sigma} -\epsilon_\k+\epsilon_\q}
    \\
    + \frac{g_{f\sigma}^2}{V} \sum_{\k}
    \frac{1}{\epsilon_{\p_2 \sigma} - \epsilon_{\p_1 \sigma} -\epsilon_\k+\epsilon_{\p_1-\p_2+\k}} 
    + \frac{g_{f\sigma}^2}{V} \sum_{\k} 
    \frac{1
    }{\epsilon_{\p_1 \sigma} - \epsilon_{\p_2 \sigma} -\epsilon_\k+\epsilon_{\p_2-\p_1+\k}}
    \\
    + \frac{g_{f\sigma}^2 g_{\sigma\sigma}}{V^2} \sum_{\K}
    \sum_{\k'\q'} \left[ \frac{(\delta_{\k'-\q',\K}  + \delta_{\q'-\k',\K + \p_1 - \p_2)}
    }{(\epsilon_{\p_2 \sigma} - \epsilon_{\p_2+\q'-\k' \sigma} -\epsilon_{\k'}+\epsilon_{\q'})(\epsilon_{\p_1 \sigma} + \epsilon_{\p_2 \sigma} -\epsilon_{\p_1+\K \sigma} - \epsilon_{\p_2-\K \sigma})} \right.\nonumber\\
    + \left. \frac{(\delta_{\q'-\k',\K} + \delta_{\k'-\q',\K + \p_1 - \p_2})
    }{(\epsilon_{\p_1 \sigma} - \epsilon_{\p_1+\q'-\k' \sigma} -\epsilon_{\k'}+\epsilon_{\q'})(\epsilon_{\p_1 \sigma} + \epsilon_{\p_2 \sigma} -\epsilon_{\p_1+\K \sigma} - \epsilon_{\p_2-\K \sigma})} \right]\; . 
\end{multline}
In this expression, we have neglected terms that vanish in the $V\to \infty$ limit. As expected, in the limit $g_{\sigma\sigma}=0$, one recovers Eq.~\eqref{eq:induced-inter_Fermi-med}. By taking the limit $\p_1,\p_2 \to 0$, we obtain
\begin{multline}
F_{n,\sigma\sigma}
= -\frac{3n}{2E_F} g_{f\sigma}^2 \\
    +  2 g_{\sigma\sigma}\left\{ 1
    + 2\frac{g_{f\sigma}^2}{V^2} \sum_{\k\q} \frac{1}{\epsilon_{\q-\k \sigma} +\epsilon_\k - \epsilon_\q}\left[\frac{1}{\epsilon_{\q-\k \sigma} +\epsilon_\k - \epsilon_\q} + \frac{1}{\epsilon_{\q-\k \sigma}}\right]\right\}\\
    = -\frac{3n}{2E_F} g_{f\sigma}^2 + 2 g_{\sigma\sigma}\left[1 + 9 \left(\frac{g_{f\sigma} n}{E_F}\right)^2B(z_\sigma )\right]\; ,
\end{multline}
where the function $B(z_\sigma )$ was defined in~\eqref{eq:B-func} and plotted in Fig.~\ref{fig:AB_functions}.

\subsection{Distinguishable impurities}
The variational state for $\sigma\neq\sigma'$ reads
\begin{multline}
    \ket{\Psi} = \left[ \alpha_{\p_1\p_2} c^{\dag}_{\p_1 \sigma} c^{\dag}_{\p_2 \sigma'} +
  \sum_{\k\q} \rho_{\p_1\p_2\k\q}
  c^{\dag}_{\p_1\sigma} f^{\dag}_{\k} f_{\q}^{} c^{\dag}_{\p_2 -\k+\q \sigma'}\right.\\
  \left. + \sum_{\k\q} \eta_{\p_1\p_2\k\q}
  c^{\dag}_{\p_1 -\k+\q \sigma} f^{\dag}_{\k} f_{\q}^{} c^{\dag}_{\p_2 \sigma'}+ \sum_{\K\neq0}\gamma_{\p_1\p_2\K}c^\dag _{\p_1+\K \sigma}c^\dag_{\p_2-\K \sigma'}  \right]\fs\; ,
\end{multline}
which satisfies the normalization:
\begin{equation}
    1 = \bra{\Psi} \ket{\Psi} =  |\alpha_{\p_1\p_2}|^2 + \sum_{\k\q} |\rho_{\p_1\p_2\k\q}|^2 +  \sum_{\k\q} |\eta_{\p_1\p_2\k\q}|^2 + \sum_{\K\neq0}|\gamma_{\p_1\p_2\K}|^2 \; .
\end{equation}
The equations of motion are given by
\begin{subequations}
\begin{align}
    E \alpha_{\p_1\p_2} =& \left(\epsilon_{\p_1 \sigma} + \epsilon_{\p_2 \sigma'} + g_{f\sigma} n + g_{f\sigma'} n + \frac{g_{\sigma\sigma'}}{V}\right)\alpha_{\p_1\p_2} + \frac{g_{f\sigma'}}{V} \sum_{\k\q} \rho_{\p_1\p_2 \k\q}\nonumber\\
    &+ \frac{g_{f\sigma}}{V} \sum_{\k\q} \eta_{\p_1\p_2 \k\q}  + \frac{g_{\sigma\sigma'}}{V} \sum_{\K} \gamma_{ \p_1 \p_2\K} \\
    E \rho_{\p_1\p_2 \k\q} =& \left(\epsilon_{\p_1 \sigma} + E_{\p_2\k\q \sigma'} + g_{f\sigma} n +  g_{f\sigma'} n + \frac{g_{\sigma\sigma'}}{V}\right)\rho_{\p_1\p_2 \k\q}  +  \frac{g_{f\sigma'}}{V} \alpha_{\p_1\p_2} + \frac{g_{f\sigma}}{V} \gamma_{\k-\q}\nonumber\\
    &+ \frac{g_{\sigma\sigma'}}{V} \eta_{\p_1\p_2 \k\q}\\
    E \eta_{\p_1\p_2 \k\q} =& \left(\epsilon_{\p_2 \sigma'} + E_{\p_1\k\q \sigma} + g_{f\sigma} n + g_{f\sigma'} n + \frac{g_{\sigma\sigma'}}{V}\right)\eta_{\p_1\p_2 \k\q}   + \frac{g_{f\sigma}}{V} \alpha_{\p_1\p_2} + \frac{g_{f\sigma'}}{V} \gamma_{\q-\k} \nonumber\\
    &+ \frac{g_{\sigma\sigma'}}{V} \rho_{\p_1\p_2 \k\q}\\
    E \gamma_{\p_1\p_2\K} =& \left(\epsilon_{\p_1 + \K \sigma} + \epsilon_{\p_2 - \K \sigma'} + g_{f\sigma} n +g_{f\sigma'} n + \frac{g_{\sigma\sigma'}}{V}\right)\gamma_{\p_1\p_2\K} + \frac{g_{\sigma\sigma'}}{V} \sum_{\K'\ne\K} \gamma_{\p_1 \p_2, \K'}\nonumber\\
    &+ \frac{g_{\sigma\sigma'}}{V} \alpha_{\p_1 \p_2} + \frac{g_{f\sigma}}{V} \sum_{\k'\q'}\rho_{\p_1 \p_2, \k'\q'}\delta_{\k'-\q',\K} + \frac{g_{f\sigma'}}{V} \sum_{\k'\q'} \eta_{\p_1 \p_2, \k'\q'}\delta_{\q'-\k',\K}\; ,
\end{align}
\end{subequations}
and can be solved perturbatively following a similar procedure as the one employed in the previous sections, leading to:
\begin{multline}
    E \simeq  E_{\mathrm{pol},\sigma} (\p_1) + E_{\mathrm{pol},\sigma'} (\p_2) +  \frac{g_{\sigma\sigma'}}{V}\\
    + 2 \frac{g_{\sigma\sigma'} g_{f\sigma} g_{f\sigma'}}{V^3} \sum_{\k\q}  \left(\frac{\frac{1}{\epsilon_{\p_1 \sigma} + \epsilon_{\p_2 \sigma'} -\epsilon_{\p_1+\k-\q \sigma} - \epsilon_{\p_2-\k+\q \sigma'}} + \frac{1}{\epsilon_{\p_1 \sigma} - \epsilon_{\p_1+\q-\k \sigma} -\epsilon_\k+\epsilon_\q}}{\epsilon_{\p_2 \sigma'} - \epsilon_{\p_2+\q-\k \sigma'} -\epsilon_\k+\epsilon_\q}\right.\\
    \left. + \frac{\frac{1}{\epsilon_{\p_1 \sigma} + \epsilon_{\p_2 \sigma'} -\epsilon_{\p_1+\q-\k \sigma} - \epsilon_{\p_2+\k -\q \sigma'}}}{\epsilon_{\p_1 \sigma} - \epsilon_{\p_1+\q-\k \sigma} -\epsilon_\k+\epsilon_\q}\right)\; .
\end{multline}
This expression recovers Eq.~\eqref{eq:FP_ind-imp_same-momenta} when $\sigma=\sigma'$ and $\p_1 = \p_2 = \p$. By taking the limits $\p_1, \p_2 \to 0$, one obtaines the following expression for the polaron interaction:
\begin{multline}
    F_{n,\sigma\sigma'} 
    = g_{\sigma\sigma'}\left\{1 + 2 \frac{g_{f\sigma} g_{f\sigma'}}{V^2} \sum_{\k\q}\left[\frac{1}{\epsilon_{\q-\k \sigma} +\epsilon_{\q-\k \sigma'}} \left(\frac{1}{\epsilon_{\q-\k \sigma} +\epsilon_\k -\epsilon_\q}\right.\right.\right.\\
    \left.\left.\left. + \frac{1}{\epsilon_{\q-\k \sigma'} +\epsilon_\k -\epsilon_\q}\right) + \frac{1}{(\epsilon_{\q-\k \sigma} +\epsilon_\k -\epsilon_\q)(\epsilon_{\q-\k \sigma'} +\epsilon_\k -\epsilon_\q)}\right]\right\}\\
    = g_{\sigma\sigma'}\left[1 + 9 g_{f\sigma} g_{f\sigma'} \left(\frac{n}{E_F}\right)^2 B(z_{\sigma},z_{\sigma'})\right]\; .
\end{multline}

\bibliography{bosemixture_long.bib}

\end{document}